\documentclass[journal]{IEEEtran}

\usepackage[dvipsnames]{xcolor}
\usepackage[per-mode=repeated-symbol,retain-explicit-plus=true]{siunitx}
\usepackage{glossaries}
\usepackage{cite}
\usepackage{graphicx}
\usepackage{booktabs}
\usepackage{multirow}
\usepackage{pgfplots}
\usepackage{subcaption}
\usepackage{pifont}
\usepackage{circledsteps}
\usepackage{hyperref}
\usepackage{cleveref}
\usepackage{orcidlink}
\usepackage{colortbl}
\usepackage{soul}
\usepackage{soulpos}
\usepackage{float}

\pgfplotsset{width=7cm,compat=1.3}

\newcommand{\cmark}{{\ding{52}}}
\newcommand{\xmark}{{\ding{56}}}

\newif\ifblind
\blindfalse

\newif\ifhighlightchanges
\highlightchangesfalse

\newcommand{\removed}[1]{\ifhighlightchanges\setstcolor{red}\st{#1}\else\ignorespaces\fi}

\newcommand{\added}[1]{\ifhighlightchanges\textcolor{blue}{#1}\else#1\fi}

\newacronym[longplural={Scratchpad Memories}]{SPM}{SPM}{Scratchpad Memory}
\newacronym[longplural={Standard Cell Memories}]{SCM}{SCM}{Standard Cell Memory}
\newacronym[longplural={Static Random-Access Memories}]{SRAM}{SRAM}{Static Random-Access Memory}
\newacronym{1RW}{1RW}{single read/write port}
\newacronym{3DIC}{3D-IC}{three-dimensional integrated circuit}
\newacronym{3R1W}{3R1W}{three read ports and one write port}
\newacronym{ACE}{ACE}{AXI Coherent Extensions}
\newacronym{AI}{AI}{Artificial Intelligence}
\newacronym{AMBA}{AMBA}{Advanced Microcontroller Bus Architecture}
\newacronym{APB}{APB}{Advanced Peripheral Bus}
\newacronym{API}{API}{Application Programming Interface}
\newacronym{ASIC}{ASIC}{Application-Specific Integrated Circuit}
\newacronym[longplural={Atomic Operations}]{ATOP}{ATOP}{Atomic Operation}
\newacronym{AVX}{AVX}{Advanced Vector Extension}
\newacronym{AXI}{AXI4}{Advanced eXtensible Interface}
\newacronym{BEOL}{BEOL}{back end of the line}
\newacronym{BLAS}{BLAS}{Basic Linear Algebra Subprograms}
\newacronym{C2C}{C2C}{Chip-to-Chip}
\newacronym{C4}{C4}{controlled collapse chip connection}
\newacronym[longplural={Core Complexes}]{CC}{CC}{Core Complex}
\newacronym{CHI}{CHI}{Coherent Hub Interface}
\newacronym{CMG}{CMG}{Core Memory Group}
\newacronym{CMOS}{CMOS}{Complementary Metal-Oxide-Semi\-con\-ductor}
\newacronym{CNN}{CNN}{Convolutional Neural Network}
\newacronym{CPU}{CPU}{Central Processing Unit}
\newacronym{CSR}{CSR}{Control and State Register}
\newacronym{CTS}{CTS}{Clock Tree Synthesis}
\newacronym{DCT}{DCT}{discrete cosine transform}
\newacronym{DDR}{DDR}{Double Data Rate}
\newacronym{DDol}{D\$}{Data Cache}
\newacronym{DLP}{DLP}{Data Level Parallelism}
\newacronym{DMA}{DMA}{Direct Memory Access}
\newacronym{DRAM}{DRAM}{Dynamic Random-Access Memory}
\newacronym{DRV}{DRV}{design rule violation}
\newacronym{DSP}{DSP}{Digital Signal Processing}
\newacronym{DUT}{DUT}{Device Under Test}
\newacronym{ECL}{ECL}{Emitter-Coupled Logic}
\newacronym{EDP}{EDP}{energy-delay product}
\newacronym{F2F}{F2F}{face-to-face}
\newacronym{FBB}{FBB}{Forward Body-Biasing}
\newacronym{FDSOI}{FD-SOI}{Fully Depleted Silicon on Insulator}
\newacronym{FEOL}{FEOL}{front end of the line}
\newacronym{FF}{FF}{Flip-Flops}
\newacronym{FMA}{FMA}{Fused Multiply-Add}
\newacronym{FO4}{FO4}{Fanout-of-4 Inverter}
\newacronym{FPGA}{FPGA}{Field-Pro\-gram\-ma\-ble Gate Array}
\newacronym{FPRF}{FPR}{Floating-Point Register File}
\newacronym{FU}{FU}{Functional Unit}
\newacronym{FPU}{FPU}{Floating Point Unit}
\newacronym{GPGPU}{GPGPU}{General-Purpose \acrlong{GPU}}
\newacronym{GPRF}{GPR}{General-Purpose Register File}
\newacronym{GPU}{GPU}{Graphics Processing Unit}
\newacronym{HBM}{HBM}{High Bandwidth Memory}
\newacronym{HDL}{HDL}{Hardware Description Language}
\newacronym{HERO}{HERO}{Heterogeneous Embedded Research Platform}
\newacronym{HPC}{HPC}{High-Performance Computing}
\newacronym{IC}{IC}{integrated circuit}
\newacronym{ICG}{ICG}{Integrated Clock Gating}
\newacronym{IDol}{I\$}{Instruction Cache}
\newacronym{ILP}{ILP}{Instruction Level Parallelism}
\newacronym{IOT}{IoT}{Internet of Things}
\newacronym{IPC}{IPC}{Instructions Per Cycle}
\newacronym{IPU}{IPU}{Integer Processing Unit}
\newacronym{ISA}{ISA}{Instruction Set Architecture}
\newacronym{LLM}{LLM}{Large Language Model}
\newacronym{LMUL}{LMUL}{Vector Length Multiplier}
\newacronym{LSU}{LSU}{Load/Store Unit}
\newacronym{LVT}{LVT}{low voltage threshold}
\newacronym{MACU}{MACU}{Multiply-Accumulate Unit}
\newacronym{MIMD}{MIMD}{Multiple Instruction, Multiple Data}
\newacronym{ML}{ML}{Machine Learning}
\newacronym{MMU}{MMU}{Memory Management Unit}
\newacronym{MUL}{MUL}{multiplier}
\newacronym{MVE}{MVE}{M-Profile Vector Extension}
\newacronym{MVL}{MVL}{maximum vector length}
\newacronym{NI}{NI}{Network Interface}
\newacronym[longplural={Networks-on-Chip}]{NOC}{NoC}{Network-on-Chip}
\newacronym{NUMA}{NUMA}{non-uniform memory access}
\newacronym{PCIe}{PCIe}{Peripheral Component Interconnect Express}
\newacronym{PC}{PC}{Program Counter}
\newacronym{PDP}{PDP}{power-delay product}
\newacronym{PE}{PE}{Processing Element}
\newacronym{PG}{PG}{Power Gating}
\newacronym{PL}{PL}{Programmable Logic}
\newacronym{PMCA}{PMCA}{Programmable Manycore Accelerator}
\newacronym{PPA}{PPA}{power, performance, and area}
\newacronym{PSL}{PSL}{Power Service Layer}
\newacronym{PTE}{PTE}{page-table entry}
\newacronym{PTW}{PTW}{page-table walker}
\newacronym{PULP}{PULP}{Parallel Ultra Low Power}
\newacronym{RAM}{RAM}{Random-Access Memory}
\newacronym{RAW}{RAW}{read-after-write}
\newacronym{RBB}{RBB}{Reverse Body-Biasing}
\newacronym{ROB}{RoB}{Reorder Buffer}
\newacronym{RR}{RR}{Round-Robin}
\newacronym{RTL}{RTL}{Register Transfer Level}
\newacronym{RV}{RV}{RISC-V}
\newacronym{RVT}{RVT}{Regular Voltage Threshold}
\newacronym{RVV}{RVV}{RISC-V Vector Extension}
\newacronym{RoCC}{RoCC}{Rocket Custom Coprocessor Interface}
\newacronym{SDRAM}{SDRAM}{synchronous dynamic random-access memory}
\newacronym{SIMD}{SIMD}{Single Instruction, Multiple Data}
\newacronym{SIMT}{SIMT}{Single Instruction, Multiple Thread}
\newacronym{SLDU}{SLDU}{Slide Unit}
\newacronym{SLVT}{SLVT}{super-low voltage threshold}
\newacronym{SM}{SM}{Streaming Multiprocessor}
\newacronym{SOA}{SoA}{State-of-the-Art}
\newacronym{SOC}{SoC}{System-on-Chip}
\newacronym{SSE}{SSE}{Streaming SIMD Extension}
\newacronym{SSR}{SSR}{Stream Semantic Register}
\newacronym{STA}{STA}{Static Timing Analysis}
\newacronym{STCO}{STCO}{System-Technology Co-Optimization}
\newacronym{SVE}{SVE}{Scalable Vector Extension}
\newacronym{TLP}{TLP}{Thread Level Parallelism}
\newacronym{TSV}{TSV}{through-silicon via}
\newacronym{TVLSI}{TVLSI}{IEEE Transactions on Very Large Scale Integration (VLSI) Systems}
\newacronym{TxnID}{TxnID}{Transaction ID}
\newacronym{VAC}{VAC}{Vector Access}
\newacronym{VAU}{VAU}{Vector Arithmetic Unit}
\newacronym{VCONV}{VCONV}{Vector Conversion}
\newacronym{VC}{VC}{virtual channel}
\newacronym{VEX}{VEX}{Vector Execute}
\newacronym{VFU}{VFU}{vector functional unit}
\newacronym{VID}{VID}{Vector Instruction Decode}
\newacronym{VIS}{VISSUE}{Vector Instruction Issue}
\newacronym{VLA}{VLA}{Vector-Length Agnostic}
\newacronym{VLSI}{VLSI}{Very Large-Scale Integration}
\newacronym{VLIW}{VLIW}{Very Long Instruction Word}
\newacronym{VLOOP}{VLOOP}{Vector Loop}
\newacronym{VLR}{VLR}{vector length register}
\newacronym{VLSU}{VLSU}{Vector Load/Store Unit}
\newacronym{VNB}{VNB}{Von Neumann Bottleneck}
\newacronym{VPU}{VPU}{Vector Processing Unit}
\newacronym{VRF}{VRF}{Vector Register File}
\newacronym{VSLDU}{VSLDU}{Vector Slide Unit}
\newacronym{VT}{VT}{vector thread}
\newacronym{W2W}{W2W}{wafer-to-wafer}
\newacronym{WAR}{WAR}{write-after-read}
\newacronym{WAW}{WAW}{write-after-write}
\newacronym{WNS}{WNS}{Worst Negative Slack}
\newacronym{XBAR}{Xbar}{Crossbar}
\newacronym{XP}{XP}{Crosspoint}

\definecolor{PULPRed}{HTML}{A8322C}
\definecolor{PULPBlue}{HTML}{1269B0}
\definecolor{PULPGreen}{HTML}{168638}
\definecolor{PULPOrange}{HTML}{F29545}
\definecolor{PULPPurple}{HTML}{910569}
\definecolor{PULPOlive}{HTML}{48592C}
\definecolor{PULPMarine}{HTML}{007996}
\definecolor{PULPGray}{HTML}{ABABAB}
\definecolor{Red}{HTML}{FF0000}

\colorlet{color1}{PULPBlue}
\colorlet{color2}{PULPRed}
\colorlet{color3}{PULPGreen}
\colorlet{color4}{PULPOrange}
\colorlet{color5}{PULPPurple}
\colorlet{color6}{PULPOlive}
\colorlet{color7}{PULPMarine}

\colorlet{colorCore}{PULPRed}
\colorlet{colorMemory}{PULPBlue}
\colorlet{colorInterconnect}{PULPGreen}
\colorlet{colorAccelerator}{PULPOrange}
\colorlet{colorPeripheral}{PULPPurple}
\colorlet{colorAlert}{Red}

\DeclareSIUnit\bank{bank}
\DeclareSIUnit\cycle{cycle}
\DeclareSIQualifier\double{DP}
\DeclareSIQualifier\single{SP}
\DeclareSIUnit\flop{FLOP}
\DeclareSIUnit\flops{FLOPS}
\DeclareSIUnit\gate{GE}
\DeclareSIUnit\op{OP}
\DeclareSIUnit\ops{OPS}
\DeclareSIUnit\bitpersecond{bps}
\DeclareSIUnit\link{link}
\DeclareSIUnit\request{request}
\DeclareSIUnit\core{core}
\DeclareSIUnit\pin{pin}
\DeclareSIUnit\hop{hop}
\DeclareSIUnit\byte{B}
\DeclareSIUnit\bit{bit}
\DeclareSIUnit\percent{\%}

\begin{document}

\bstctlcite{IEEE:BSTcontrol}

\title{FlooNoC: A \SI{645}{\giga\bitpersecond\per\link} \SI{0.15}{\pico\joule\per\byte\per\hop} Open-Source \gls{NOC} with Wide Physical Links and End-to-End \gls{AXI} Parallel Multi-Stream Support}

\author{Tim~Fischer\orcidlink{0009-0007-9700-1286},~\IEEEmembership{Student Member, IEEE},
        Michael~Rogenmoser\orcidlink{0000-0003-4622-4862},~\IEEEmembership{Student Member, IEEE},
        Thomas~Benz\orcidlink{0000-0002-0326-9676},~\IEEEmembership{Student Member, IEEE},
        Frank~K.~Gürkaynak\orcidlink{0000-0002-8476-554X},
        and~Luca~Benini\orcidlink{0000-0001-8068-3806},~\IEEEmembership{Fellow, IEEE}
\thanks{T. Fischer, M. Rogenmoser, T. Benz, F. K. Gürkaynak, and L. Benini are with
the Integrated System Laboratory (IIS), ETH Zurich, Switzerland. E-mail:
{fischeti, michaero, tbenz, kgf, lbenini}@iis.ee.ethz.ch}
\thanks{L. Benini is also with the Department of Electrical, Electronic and Information Engineering (DEI), University of Bologna, Italy.}
\thanks{This work has been supported in part by ‘The European Pilot’ project under grant agreement No 101034126 that receives funding from The European High Performance Computing Joint Undertaking (EuroHPC JU) as part of the EU Horizon 2020 research and innovation programme.}
\thanks{©2025 IEEE. Personal use of this material is permitted. Permission from IEEE must be obtained for all other uses, in any current or future media, including reprinting/republishing this material for advertising or promotional purposes, creating new collective works, for resale or redistribution to servers or lists, or reuse of any copyrighted component of this work in other works.}
\thanks{IEEE Publication DOI: 10.1109/TVLSI.2025.3527225}}

\markboth{IEEE Transactions on Very Large Scale Integration (VLSI) Systems,~VOL.~(33), NO.~(4), APRIL 2025}%
{Shell \MakeLowercase{\textit{et al.}}: Bare Demo of IEEEtran.cls for IEEE Journals}
%



\maketitle

\begin{abstract}

The new generation of domain-specific AI accelerators is characterized by rapidly increasing demands for bulk data transfers, as opposed to small, latency-critical cache line transfers typical of traditional cache-coherent systems. In this paper, we address this critical need by introducing the \textit{FlooNoC} Network-on-Chip (NoC), featuring very wide, fully Advanced eXtensible Interface (AXI4) compliant links designed to meet the massive bandwidth needs at high energy efficiency.
At the transport level, non-blocking transactions are supported for latency tolerance. Additionally, a novel end-to-end ordering approach for AXI4, enabled by a multi-stream capable Direct Memory Access (DMA) engine simplifies network interfaces and eliminates inter-stream dependencies. 

Furthermore, dedicated physical links are instantiated for short, latency-critical messages. A complete end-to-end reference implementation in 12nm FinFET technology demonstrates the physical feasibility and power performance area (PPA) benefits of our approach. Utilizing wide links on high levels of metal, we achieve a bandwidth of \SI{645}{\giga\bitpersecond} per link and a total aggregate bandwidth of \SI{103}{\tera\bitpersecond} for an 8$\times$4 mesh of processors cluster tiles, with a total of 288 RISC-V cores. The NoC imposes a minimal area overhead of only 3.5\% per compute tile and achieves a leading-edge energy efficiency of \SI{0.15}{\pico\joule\per\byte\per\hop} at \SI{0.8}{\volt}. Compared to state-of-the-art NoCs, our system offers three times the energy efficiency and more than double the link bandwidth. Furthermore, compared to a traditional AXI4-based multi-layer interconnect, our NoC achieves a 30\% reduction in area, corresponding to a 47\% increase in \SI{}{\giga\flops\double} within the same floorplan.

\end{abstract}


%
\IEEEpeerreviewmaketitle

\section{Introduction}
%

The demands of modern workloads, particularly \glspl{LLM}, have led to the emergence of ultra-large \gls{AI} workload accelerators fabricated in full reticles of cutting-edge technologies~\cite{intel_gaudi_hc24, meta_mtia_hc24, microsoft_maia_hc24}. These accelerators feature tile-based designs arranged in tiled floorplans and connected with mesh \glspl{NOC}. Due to the substantial bandwidth requirements, these systems commonly rely on \glspl{NOC} designed to support bulk data transfers. Moreover, the memory-intensive nature of these workloads results in massive bandwidth at the boundary of the mesh, as the tiles frequently require expensive off-chip memory accesses~\cite{memory_wall, squeezeLLMmemorybound}. Addressing these issues is crucial, requiring \glspl{NOC} that are designed for both high bandwidth and latency tolerance to enable asynchronous data processing while maximizing memory bandwidth utilization.


In this paper, we focus on \glspl{NOC} for the new wave of extreme AI accelerators, optimized for bulk data transfers, typically generated by \gls{DMA} engines. Unlike \glspl{NOC} for cache-coherent systems that are designed to transport cache lines at low latency, bulk-transfer networks are designed to handle large chunks of data efficiently with an emphasis on bandwidth rather than latency. Cache-coherent systems and interconnects, which have also been the subject of intensive investigation, are out-of-scope for this work; we refer the interested reader to existing literature on this topic \cite{cache_coherency_noc, no_wide_flits}.

\begin{figure}[ht]
    \centering
    \includegraphics[width=\columnwidth]{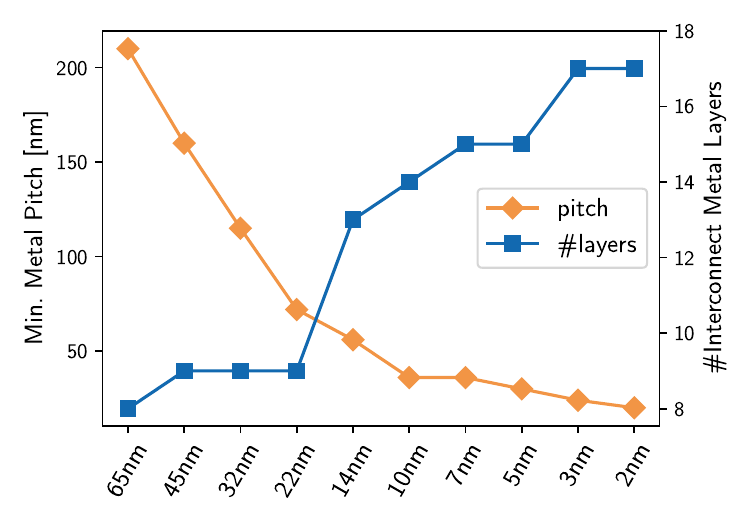}
    \caption{Technology scaling of on-chip wire resources based on IDRS reports \cite{IRDS} for 2-\SI{14}{\nano\meter} and \cite{interco_scaling16} for 22-\SI{65}{\nano\meter}.}
    \label{fig:tech_scaling}
\end{figure}

Delivering large bulks of data to the \glspl{PE} with a sustained high-bandwidth data flow is an unsolved problem at the scale needed today and in the near future. In traditional \glspl{NOC}, large data transfers are serialized over many flits onto links up to \SI{64}{\bit} in width~\cite{jerger2022chip}, with increased link frequency to satisfy bandwidth requirements. However, this approach encounters significant limitations in modern systems, where bandwidth is ultimately constrained by the on-chip clock frequency, which cannot be raised above a few \SI{}{\giga\hertz} for power and signal integrity reasons. Consequently, the traditional practice of serialized links requires reassessment in modern \glspl{NOC}. 

Furthermore, as illustrated in \Cref{fig:tech_scaling}, recent advancements in on-chip routing resources reveal that within the past decade, the minimum metal pitch in \gls{VLSI} technologies has decreased tenfold while the number of interconnect metal layers has more than doubled. This trend strongly supports a shift towards wider links as opposed to faster links to accommodate the ever-increasing demand for bandwidth.

The trend in technology scaling shown in \Cref{fig:tech_scaling} also calls for a re-evaluation of the use of virtual channels in \gls{NOC} designs. Traditionally, \glspl{NOC} have employed virtual channels to enable sharing and the reduction of physical links. However, with the advancements in modern \gls{VLSI} technologies, which now provide extensive routing resources sufficient for multiple and wide physical links, the justification for using virtual channels is diminishing. Although virtual channels can enhance network routability by reducing the number of wires, they also affect bandwidth efficiency and add complexity to routers due to the need for additional buffering and arbitration between virtual channels. Given these factors, a shift toward physical links over virtual ones is increasingly justified in today's \gls{VLSI} technologies~\cite{dally_keynote}.

Although routing resources in newer technologies are becoming increasingly abundant, merely increasing the width of \gls{NOC} links is not sufficient for an efficient implementation. Integrating the \gls{NOC} tightly with logic blocks results in better area utilization compared to their physical separation~\cite{batten_noc_symbiosis}. However, this integration poses additional challenges, such as potential impacts on compute logic due to the increased routing resource demands of the \gls{NOC}. Furthermore, \gls{NOC} designers must consider the physical distances that links must span, adding appropriate buffer cells or elastic buffers to achieve targeted timing~\cite{dimitrakopoulos2015microarchitecture}. This task is complicated by modern systems incorporating substantial on-chip memory in the form of \gls{SRAM} macros, which can obstruct buffer placement and block routing on the lower metal layers. Consequently, a deep understanding of the underlying technology and synergy with the physical design flow is essential to navigate these complexities effectively during the \gls{NOC} design process.

In addition to addressing physical constraints, one must consider how data transfers are managed at the transport level. Non-coherent block data transfers, which require explicit management, typically via \gls{DMA} engines, provide an efficient way to control data movement on-chip and towards main memory. \gls{AXI}, as the prevailing protocol for non-coherent transfers, is particularly suited for these tasks. It supports burst transfers and multiple outstanding transactions, enhancing its latency tolerance and making it a preferred choice for non-coherent systems. Despite its advantages and widespread adoption, implementing \gls{AXI} as a network-level protocol in a \gls{NOC} introduces significant challenges. Its stringent requirement for transaction ordering based on the \gls{TxnID} complicates operations, as the complexity of monitoring outstanding transactions and maintaining order scales with the network’s diameter~\cite{kurth_axi}. This complexity significantly restricts the scalability of multi-hop interconnects that depend on \gls{AXI}-compliant links.

Separating the \gls{NOC} network-level protocol from the \gls{AXI}-compliant initiators with \glspl{NI}~\cite{benini_og_noc, jerger2022chip} addresses the inherent scalability limitations of \gls{AXI} and enables the design of more flexible and scalable interconnect architectures suited for many-core systems. Specifically, implementing a bespoke \gls{NOC} network-level protocol simplifies the router design significantly. This setup enables packets to be routed based primarily on their destination addresses while the \glspl{NI} efficiently handle outstanding transactions and guarantee the ordering of responses. 
While there has been a surge in open-source \gls{NOC} implementations \cite{esp_noc_soc, celerity, openpiton}, the availability of \gls{AXI}-compliant \glspl{NI} that support multiple transactions and burst transfers remain confined to proprietary industry solutions \cite{arm_noc, flexnoc, signatureip_nc_noc}, whose performance and scalability are not assessed and quantified in the open literature.

A common strategy to implement a high-performance, \gls{AXI}-compliant \gls{NI} that adheres to \gls{AXI}’s ordering restrictions is to use a \gls{ROB} to manage \gls{AXI} responses that arrive out of order from the \gls{NOC}~\cite{floonoc, axi_ni_rob, ni_rob}. However, this solution presents several significant disadvantages. First, \glspl{ROB} are costly due to their large memory requirements. For example, with a single \gls{AXI} burst capable of reaching up to \SI{4}{\kilo\byte}, the \gls{ROB} must be sized to accommodate several such bursts to manage multiple outstanding transactions. Second, the performance of the \gls{NI} is closely tied to the \gls{ROB}’s capacity; transaction processing must be stalled if the \gls{ROB} is full. Moreover, while \glspl{ROB} enable the handling of multiple out-of-order transactions, they inherently increase response latency, as each response might be delayed in the buffer before it is forwarded to the \gls{AXI} interface. Consequently, a more holistic end-to-end approach, rather than a strictly \gls{NI}-centric solution, is necessary to address these challenges effectively.

Given all these issues that arise in modern \glspl{NOC} for high-performance systems, we propose a new \gls{NOC} design based on the following four principles: \textbf{1) Wide links} routed on high metal levels, supporting buffer insertion in the design implementation to mitigate the serialization bottleneck and substantially increasing bandwidth to accommodate data-intensive applications. \textbf{2) Physical channels over virtual ones}, enhancing the network’s ability to manage diverse traffic types effectively. \textbf{3) Full \gls{AXI} support}, enabling multiple outstanding transactions and burst capabilities to meet high-performance requirements efficiently. \textbf{4) Decoupling network and transport layers} to solve the scalability issues of \gls{AXI} multi-hop networks.

Compared to our preliminary work~\cite{floonoc}, we have incorporated several critical improvements to the architecture and present new results. First, we have expanded the \gls{NI} by introducing a \textit{\gls{ROB}-less} implementation, significantly enhancing area efficiency. Second, we have integrated a multi-stream capable \gls{DMA} engine into the compute cluster, which enables end-to-end ordering more effectively than our previous work. Third, we substantially improved the physical implementation of the \gls{NOC}, integrating it with a compute cluster into a compute tile, achieving significant advancements across \gls{PPA} metrics. Furthermore, we have carried out physical implementation on a complete large-scale design comprising 288 RISC-V cores and conducted a detailed \gls{PPA} analysis for the entire system. Lastly, we carried out a quantitative comparison between \textit{FlooNoC} and an interconnect based on multi-level \gls{AXI}-\glspl{XBAR} design,  demonstrating our solution's superior performance and area benefits. The key contributions of this paper are:

\begin{itemize}
    \item We present (to the best of our knowledge) the first open-source\footnote{https://github.com/pulp-platform/FlooNoC} \gls{NOC} with fully \gls{AXI}-compliant initiator and target interfaces that efficiently handle the ordering requirements imposed by \gls{AXI} at the endpoints rather than in the routers while achieving full bandwidth utilization.
    \item We propose an end-to-end ordering solution that combines a \gls{ROB}-less implementation of the \gls{NI} with a multi-stream capable \gls{DMA}. The tight integration thereof in a compute tile eliminates inter-stream dependencies, offering a streamlined, high-performance solution while reducing area complexity by up to \removed{58\%}\added{\SI{256}{\kilo\gate}}.
    \item We demonstrate the physical implementation of an $8 \times 4$ compute mesh using \SI{12}{\nano\meter} \gls{VLSI} technology. The \gls{NOC} integrated into the compute tile accounts for a mere 3.5\% of the tile area, yet achieves leading-edge energy-efficiency and performance, delivering \SI{0.15}{\pico\joule\per\byte\per\hop} at \SI{0.8}{\volt} and an aggregate bandwidth of \SI{103}{\tera\bitpersecond} at \SI{1.26}{\giga\hertz} in typical conditions, corresponding to a delay of less than 70 \gls{FO4}.
    \item We present a comprehensive comparison between a \gls{SOC} designed with traditional \gls{AXI} matrices and our \gls{NOC}-based solution. Our findings demonstrate that the \gls{NOC}-based solution achieves a 30\% reduction in area for the same core count or delivers 47\% higher performance given the same floorplan.
\end{itemize}

\section{Background \& Related Works}

Research in \glspl{NOC} has significantly advanced since its inception \cite{benini_og_noc, dally_og_noc} more than two decades ago, yet many areas remain actively explored. We break down our discussion of related works into three main areas of \gls{NOC} research tackled by our study. First, we look at efforts to ensure compatibility with the widely-used \gls{AXI} standard and its ordering rules, a critical aspect for interfacing with existing systems. Then, we move on to the development of wide and physical channels essential for high-bandwidth systems. Lastly, we consider the physical awareness of \gls{NOC} design, stressing the importance of making design choices that align with the physical constraints of \gls{VLSI} design.

\subsection{Advanced Microcontroller Bus Architecture (AMBA)}

The \gls{AMBA} is a family of open on-chip protocol specifications developed by ARM that have continuously evolved to meet the demands of modern systems. \gls{AXI} is a widely-used protocol for high-bandwidth on-chip communication~\cite{axi_spec} with several well-established open-source implementations~\cite{kurth_axi}. \Gls{AXI} defines separate channels for read and write requests (AR, AW, W) and response channels (R, B). It also supports multiple outstanding transactions, which allows memory latency to be hidden at the initiator for a higher system throughput.

The \gls{AXI} protocol enforces strict ordering of transactions using the \gls{TxnID}, which serves as an identifier for each transaction. The width of the \gls{TxnID} is determined by the number of initiators and their characteristics. The protocol requires that transactions with the same \gls{TxnID} are processed in order and, due to its role in routing responses, that the \gls{TxnID} width increases with each network hop to maintain unique transaction identifiers. This requirement creates challenges in scaling \gls{AXI} for large systems with multiple hops, as it requires managing state information for each \gls{TxnID}, leading to exponential complexity~\cite{kurth_axi} and limiting scalability. While \gls{TxnID} remappers~\cite{kurth_axi} can reduce the number of IDs, they introduce significant overhead in latency and area and complicate tracing and verification.

The latest generation of the \gls{AMBA} family, \gls{AMBA}5, introduced the \gls{CHI} protocol, designed for high-performance systems requiring cache coherency. \gls{CHI} operates at a fine granularity, handling data at the cache line level, and is not optimized for bulk data transfers. In contrast, \gls{CHI} complements, rather than replaces, \gls{AXI}, which is further evidenced by the simultaneous introduction of AXI5, which features performance improvements and enhanced functionality to align with \gls{CHI}’s capabilities. Among them, AXI5 introduced support for \glspl{ATOP}. The main difference between atomic and non-atomic transactions is that the former generates both a read and a write response on the \textit{R}, respectively, \textit{B} channel. To prevent issues related to the ordering of transactions, the \gls{AXI} specification requires that outstanding atomic transactions have a unique \gls{TxnID}. Furthermore, atomic transactions cannot use the same \gls{TxnID} as non-atomic transactions that are outstanding.

This work is fully compatible with the \gls{AXI} standard, leveraging its support for multiple outstanding transactions and burst transfer capabilities to achieve high performance. Additionally, we incorporate support for \glspl{ATOP}, introduced with AXI5, which are essential in manycore systems for data consistency and efficient synchronization of \glspl{PE}.

\subsection{AXI4 Network-on-Chips}
Previous efforts have explored leveraging \gls{AXI} matrices to develop an \gls{AXI}-compatible \gls{NOC}\cite{kurth_axi, jain2023patronoc}, using \gls{AXI}-\glspl{XP} in place of traditional routers. Such an \gls{AXI}-\gls{XP} includes a standard \gls{AXI} \gls{XBAR} equipped with \gls{TxnID} remappers at each output port, ensuring \gls{TxnID} width remains constant across the network. However, this method faces several challenges. Each \gls{XP} incurs substantial logic overhead due to the necessity of tracking outstanding transactions through the crossbar and \gls{TxnID} remappers. Additionally, maintaining transaction ordering for those with identical IDs in \gls{AXI}-\glspl{XP} can degrade performance, as it may force new transactions to stall to preserve order.

To address these ordering challenges, \glspl{ROB} have been suggested to be implemented either at endpoints\cite{axi_ni_rob, ni_rob, wang2020efficient} or within the \gls{NOC} itself\cite{router_rob}. These solutions include \gls{ROB} optimizations such as dynamic allocation and tracking entries in "ordering lists" to optimize \gls{ROB} use and minimize its size\cite{axi_ni_rob, wang2020efficient}. \gls{TxnID} renaming has also been introduced\cite{ni_rob} to manage multiple in-order transactions. Nevertheless, these strategies introduce substantial complexity. They necessitate tables to monitor \gls{ROB} entries and face limitations due to the potential size of the \gls{ROB}, which is particularly problematic in high-bandwidth systems. For example, \gls{AXI} supports data widths up to \SI{1024}{\bit} and burst sizes up to \SI{4}{\kilo\byte}. Although \glspl{ROB} may suffice for smaller transactions, they prove to be excessively costly for handling multiple outstanding bursts on the scale of kilobytes.

\subsection{Wide \& Physical Channels}
In the field of \gls{NOC} design for high bandwidth systems, recent research \cite{esp_isscc24}, as well as commercial-grade chips \cite{northpole_ibm_isscc24}, lean towards employing physical channels over virtual ones. Studies referenced in \cite{vc_vs_phys_carloni13, vc_vs_phys_bertozzi} illustrate that opting for physical channels contributes to both area and power efficiency improvements, which will gain further relevance with the progress in semiconductor technologies, where an abundance of routing resources favors simpler microarchitectures. Routers of physical channel-based \glspl{NOC} are characterized by reduced buffering requirements and the capability to operate at higher frequencies, thereby boosting overall system performance. Furthermore, the perspective offered in Bill Dally's 2020 \gls{NOC} symposium keynote\cite{dally_keynote} underscores the significance of leveraging technological advancements in routing resources, advocating for a shift towards physical channels.

Besides using physical channels over virtual ones, \textit{Ruche} networks \cite{ruche_nocs20} have been proposed, introducing additional long-range physical channels that bypass routers in 2D-mesh networks with the goal of decreasing the \gls{NOC} diameter and bisection bandwidth. They incur a significant cost in terms of area, however, since the router radix increases. Further, in scenarios where traffic is primarily routed towards I/O or memory located at the boundary of a chip, \textit{Ruche} channels will not be able to provide additional bandwidth since traffic will be predominantly routed through \textit{Ruche} channels, resulting in lower utilization of the local links.

Splitting a single physical link into multiple subnets, rather than using multiple physical links, has been proposed as a strategy to enhance energy efficiency \cite{tvlsi_bwexp}. This approach allows for more fine-grained \gls{PG} of the subnets, improving energy management. However, while this method conserves power, it also reduces the bandwidth of the links and introduces latency overheads stemming from additional serialization requirements and the time needed to wake up the subnets.

Wide physical links to achieve higher throughput are common in many commercial products \cite{northpole_ibm_isscc24, tesla_dojo_isscc23} with channel widths up to \SI{1024}{\bit}. Another study has also reported significant performance gains in \gls{RTL} simulation by increasing the link size to \SI{512}{\bit}\cite{openpiton_hpc}, but its implementation was limited to \glspl{FPGA}, and actual physical implementation was not demonstrated. There has also been research arguing for smaller flit sizes \cite{no_wide_flits}. However, the authors limited the analysis to virtual channel routers and traffic of coherent general-purpose \glspl{CPU}. Coherent traffic is very different from burst-based non-coherent traffic since it is characterized by a high amount of small control packets. Furthermore, the common assumption the authors use that the router area grows quadratically with the channel width has been disputed if physical implementation effects are considered \cite{batten_noc_symbiosis}.

\added{\gls{DDR} links have also been proposed~\cite{psarras2016ddrlink1} to reduce power consumption by minimizing wire capacitance through interleaved routing. However, they rely on router duplication to support data transfers on both clock edges. In physical-channel-based designs, this duplication introduces significant area overhead, as the typical mitigation strategy of reducing virtual channels~\cite{psarras2016ddrlink2} does not apply.}

\subsection{Awareness of Physical Implementation}
The physical implementation aspect of \glspl{NOC} has also been studied in the literature. In \cite{batten_noc_symbiosis}, the authors analyze the area and wiring resources of \gls{NOC} routers and conclude that routers are routing-bound. They propose a \textit{\gls{NOC} Symbiosis} to absorb the wiring resources of the node logic, which are typically underutilized. This approach is also evident in industrial solutions with a tile-based design \cite{tesla_dojo_isscc23, northpole_ibm_isscc24}, where the \gls{NOC} is flattened into the tile instead of implemented as a standalone block.

Furthermore, the choice of topology plays a crucial role in the physical design of \glspl{NOC}. The 2D-mesh, favored for its simplicity and efficiency, is predominantly used in both academic \cite{piton_micro17, esp_isscc24, tvlsi_cim_mesh, tvlsi_mesh_deploy, tvlsi_hipu_mesh} and commercial implementations \cite{tesla_dojo_isscc23, northpole_ibm_isscc24, cerebras_micro23, untether_ai_hc22}. \removed{However, }Exploring tile-based physical design methodologies has not been confined to \removed{conventional topologies}\added{the mesh topology}. For instance, \textit{Ruche} and torus topologies, which reduce network diameter by instantiating links spanning more than one tile on the physical layout of the die, have been \removed{evaluated}\added{proposed and analyzed in} \cite{batten_pd_nocs20}. \added{However, they create significant challenges on timing closure due to the larger physical distances spanned by their longer links, requiring elastic buffers, which would increase latency, potentially reducing the benefits of diameter reduction on latency.}

A quantitative and comparative \gls{PPA} analysis of various mesh-based \gls{SOA} \gls{NOC} physical implementations will be presented in \Cref{tab:soa_comparison}. This analysis will explore the trade-offs between achieved bandwidth and the costs in terms of area, as well as the energy efficiency of data transfers, which are crucial considerations in today’s \gls{HPC} systems.
\section{NoC Microarchitecture}

\subsection{Network Interface (NI)}
    
\begin{figure*}[ht!]
    \centering
    \includegraphics[width=\textwidth]{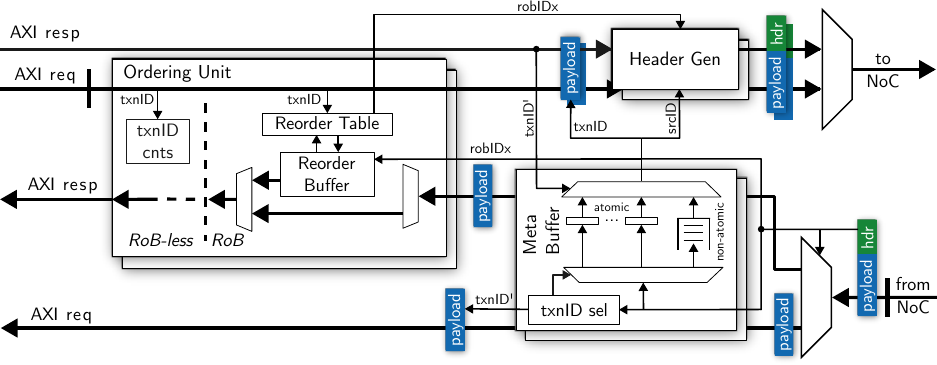}
    \caption{\gls{AXI} Network Interface architecture for request (AR/AW/W) and response (R/B) paths. Reads and writes are independent in \gls{AXI}, and the request/response paths are very similar in the \gls{NI} and are depicted as overlapping modules. The \textit{Ordering Unit} can be configured with or without reordering capabilities (i.e. \textit{RoB} and \textit{RoB-less}).}
    \label{fig:network-interface}
\end{figure*}

The main task of the \gls{NI} is to handle the protocol conversion at the endpoint from \gls{AXI} to the network-level protocol while adhering to the ordering requirements of \gls{AXI}. \gls{AXI} requires that transactions with the same \gls{TxnID} are ordered. These ordering guarantees can be satisfied using additional hardware in every switch to track outstanding transactions for each \gls{TxnID}~\cite{kurth_axi}. However, this results in significant complexity in every hop of the network, drastically limiting scalability. To avoid this complexity, we handle reordering in the \gls{NI}, allowing us to use simpler \gls{NOC} routers that do not need to guarantee in-order transactions. The architecture of the \gls{NI} is shown in \Cref{fig:network-interface}, which includes an \textit{Ordering Unit}\removed{, as shown in orange}.

The \textit{Ordering Unit} ensures the correct order of \gls{AXI} responses returning to the \gls{AXI} interface. Ordering can be achieved in two different ways: 1) buffering out-of-order responses in a \gls{ROB} and releasing them in order again, and 2) stalling the injection of new requests into the \gls{NOC} if the responses might arrive out of order, which can happen if two requests target different destinations. Our proposed \gls{NI} supports both approaches with two different configurations.

\subsubsection{Reorder-Capable NI}
The reorder-capable \gls{NI}, already outlined in~\cite{floonoc}, features a \gls{ROB} to enable concurrent outstanding transactions to different destinations without violating the ordering of the responses. It does so by keeping track of outstanding transactions in a \textit{Reorder Table} and using end-to-end flow control, meaning new \gls{AXI} requests are only injected into the \gls{NOC} if there is enough space in the \gls{ROB} to store the response. A response always contains a unique identifier \textit{robIDx}, which is stored in the \textit{Reorder Table} to determine whether it is out of order and needs to be buffered in the \gls{ROB}. Otherwise, the response can be directly forwarded to the \gls{AXI} interface. Apart from being a unique identifier, the \textit{robIDx} also acts as the index into the \gls{ROB}, where it should be stored. The space in the \gls{ROB} is allocated when the request is granted and before injecting it into the \gls{NOC}. This \gls{ROB} allocation happens dynamically, and the generated \textit{robIDx} is pushed into the \textit{Reorder Table}, where it is removed once the response has been forwarded to the \gls{AXI} interface, either from the \gls{ROB} or directly from the \gls{NOC}. We also implemented two optimizations that reduce the need for \gls{ROB} storage allocation 1) The first response of a stream of transactions with equal \gls{TxnID} is always in order and does not require allocation 2) Assuming deterministic routing in the network, the responses of requests to the same destination will arrive in the same order as the requests were issued. Hence, there is no need to reorder them. Those optimizations can be done independently for each \gls{TxnID}, which allows for the retention of support for out-of-order transactions of different \glspl{TxnID}.

\subsubsection{RoB-less NI}
We also propose a \textit{\gls{ROB}-less} \gls{NI}, a more cost-effective implementation to guarantee the ordering requirements. For each \gls{TxnID}, a counter tracks the number of outstanding transactions and stalls incoming requests if the \textit{dstID} differs from the previous transactions that are currently outstanding. Assuming a static routing algorithm inside the network, this mechanism already solves the ordering requirements since responses from the same destination are guaranteed to arrive in the same order as the corresponding requests, while responses from different destinations might arrive out of order. The lack of buffering resources makes this solution very cost-effective at the expense of potential performance degradation due to stalls. However, stalls can be prevented if transactions to different destinations are decoupled downstream with different \glspl{TxnID}.

\subsubsection{Non-Atomic \& Atomic Transactions}

On top of \gls{AXI}, which only defines non-atomic transactions, we additionally support \glspl{ATOP} as outlined in \cite{kurth_axi}. The main difference between \glspl{ATOP} and non-atomic transactions is that each \gls{ATOP} has a unique \gls{TxnID} amongst all outstanding transactions. This requirement has some implications for handling \glspl{ATOP} in the \gls{NI}. First, \glspl{ATOP} can bypass the \textit{Ordering Unit} since it is guaranteed that there are no other outstanding transactions with the same \gls{TxnID}. Second, \glspl{ATOP} are treated differently by the \textit{Meta Buffer}, which stores the information required to return the responses to the source (i.e., the \textit{srcID}) in a FIFO. Non-atomic transactions are all mapped to the same \gls{TxnID} to guarantee ordering downstream, and the \textit{srcID} for the response can be popped from the FIFO in the same order it was pushed into. \glspl{ATOP}, however, are allowed to arrive out-of-order. Hence, the \textit{Meta Buffer} has a separate set of buffers to store the return information of \glspl{ATOP}.

\subsection{Links \& Flits}
\label{subsec:link_flits}

\begin{figure}[htbp]
    \centering
    \includegraphics[width=\columnwidth]{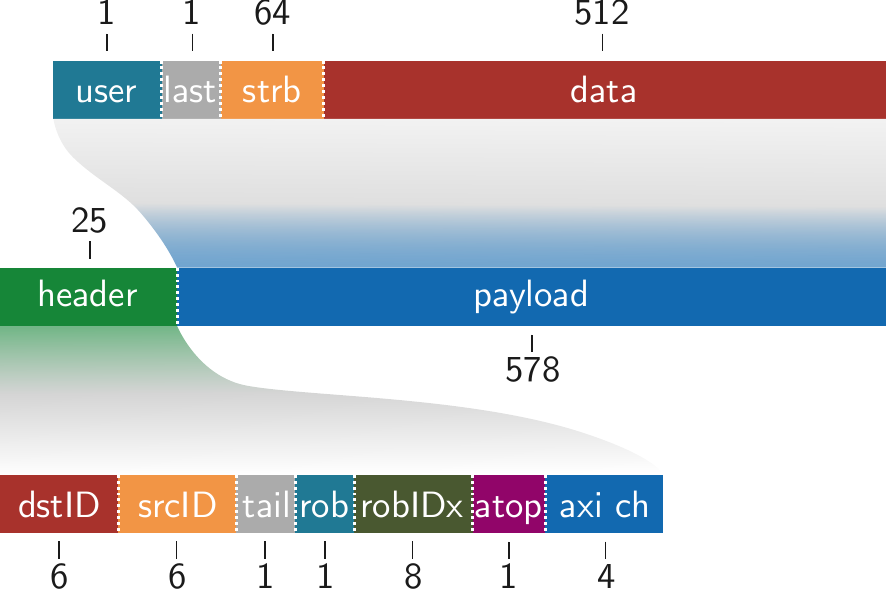}
    \caption{Example of a single flit, consisting of header information and an AXI W beat payload of \SI{512}{\bit}.}
    \label{fig:packet}
\end{figure}

The conventional method of serializing a packet over a narrow link and identifying the start and end with header and tail flits is inefficient for wide physical links. When an entire packet can be transmitted in a single flit on a wide link, the additional header and tail flits reduce the effective bandwidth to 33\%. We employ parallel lines for header information, including routing, ordering, and payload type, to overcome this limitation, as illustrated in \Cref{fig:packet}.


Besides using parallel lines for header information, we also employ separate physical channels to accommodate the diverse traffic requirements within an \gls{SOC} interconnect. The traffic can vary greatly due to the different types of initiators involved. For instance, programmable \glspl{DMA} often utilize very wide buses and burst-based data transfers to meet the high bandwidth demands of compute elements. In contrast, compute cores usually generate single-word transfers for tasks such as synchronization and configuration. To cater to these different traffic patterns, we provide dedicated links for latency-sensitive traffic via narrow channels and high-bandwidth traffic via wide channels. The dimensions of both narrow (64-bit) and wide (512-bit) links are configured to fit all packets into a single flit, which can be transmitted in one cycle. This setup enables us to match the frequency and bandwidth of the endpoint \gls{AXI} bus with the \gls{NOC} links. We implement three physical links in each direction, as \Cref{tab:channels} details.


\begin{table}[htbp]
    \centering
    \caption{Description and dimensions of physical links. Mapping of AXI requests and responses of \textsc{DataWidth} = 64/\SI{512}{\bit}, \textsc{AddrWidth} = \SI{48}{\bit}.}
    \begin{tabular}{@{}lccc@{}}
    \toprule
    & & \multicolumn{2}{c}{\textbf{Mapping \& Primary Payload}} \\
    \cmidrule(l){3-4}
    \textbf{Phys. link} & \textbf{Size [bits]} & \textbf{AXI Narrow} & \textbf{AXI Wide} \\
    \midrule
    \multirow{2}{*}{$\mathtt{req}$} & \multirow{2}{*}{119} & AR/AW: 48-bit addr & \multirow{2}{*}{AR: 48-bit addr} \\
    & & W: 64-bit data \\
    \midrule
    \multirow{2}{*}{$\mathtt{rsp}$} & \multirow{2}{*}{103} & R: 64-bit data & \multirow{2}{*}{B: 2-bit resp}\\
    & & B: 2-bit resp &  \\
    \midrule
    \multirow{2}{*}{$\mathtt{wide}$} & \multirow{2}{*}{603} & \multirow{2}{*}{-} & AW: 48-bit address \\
    & & & R/W: 512-bit data \\
    \bottomrule
    \\
    \end{tabular}
    \label{tab:channels}
\end{table}


\gls{AXI} requests and responses are always transmitted over separate physical links to avoid message-level deadlocks. The $\mathtt{req}$ and $\mathtt{rsp}$ links are mainly used to handle latency-sensitive requests and responses from compute cores. In addition, narrow links are utilized for read requests and write responses from the wide \gls{AXI} bus, as these messages do not fully utilize the bandwidth of a wide link. By mapping them to narrow links, the wide link is reserved for high-bandwidth traffic, such as read and write bursts.

An exception to this approach involves wide write requests, which are mapped to the wide link. Write beats (\textit{W} beats) are not associated with a \gls{TxnID}, so they must be strictly ordered. This ordering requirement becomes challenging when address write (\textit{AW}) and write data (\textit{W}) are sent over different links and could be interleaved with other write requests from different initiators. To address this issue, write requests and write data are always bundled together, and wormhole routing is used in the \gls{NOC} to prevent interleaving and maintain order.

\subsection{Router}

\begin{figure}[htbp]
    \centering
    \includegraphics[width=0.9\columnwidth]{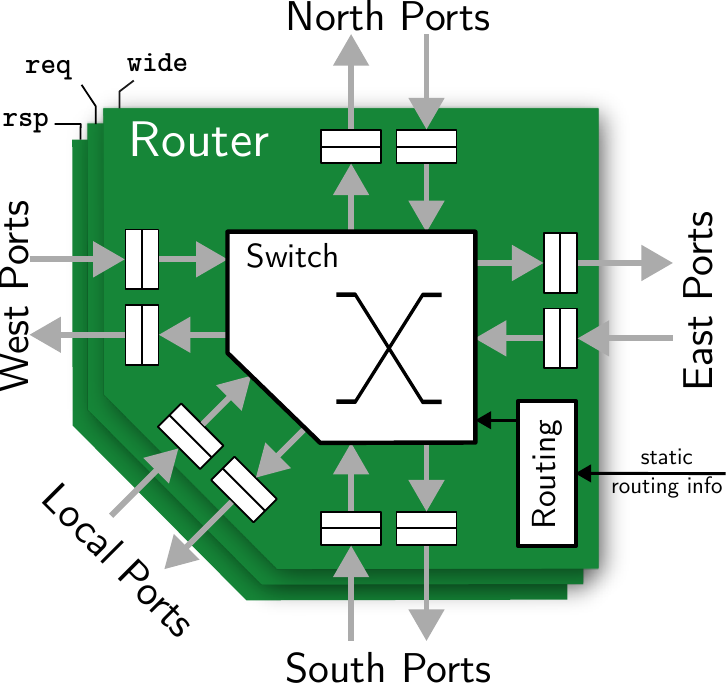}
    \caption{Router microarchitecture of a 5$\times$5 configuration with three links. Routing information such as local coordinates and routing tables are passed from externally.}
    \label{fig:router}
\end{figure}

Our \gls{NOC} leverages wide physical links and a reduced operating frequency compared to traditional narrow links, offering significant microarchitectural advantages in router design, as illustrated in \Cref{fig:router}. We utilize simple, low-area, and low-complexity routers that do not require internal pipelining. Furthermore, instead of virtual channels, we implement multilink routers, which include separate routers for each of the three physical links, ensuring complete network isolation. The routers also do not enforce any ordering of flits, which significantly 
improves scalability compared to interconnects based on \gls{AXI} matrices.

These routers are highly configurable, supporting any number of input and output ports, and use a valid-ready handshake mechanism for control flow. The routers are equipped with minimal input buffers and can optionally include output buffers, allowing a trade-off between reduced latency and improved timing closure for long routing channels \cite{dimitrakopoulos2015microarchitecture}. Additionally, the internal switch of the router is optimized to disable loopbacks and exclude impossible connections under XY-Routing. \added{Arbitration in the router switch is handled using a round-robin scheme, implemented as a logarithmic tree to ensure fairness.}


The router supports wormhole routing, which can be enabled on a flit basis. For instance, \textit{W} bursts require wormhole routing to prevent interleaving with different streams of flits. The routing decision is handled by the \textit{Routing} module, which was designed in a way that makes it easily extendable to a wide variety of routing algorithms. Currently, the router supports multiple static routing algorithms 1) \textit{Source-based} routing, where the route through the network is computed at the source and encoded in the header of the flit 2) \textit{Dimension-ordered-routing} which is applicable in 2D meshes 3) \textit{Table-based} routing, where the output port for each \textit{dstID} is stored in a routing table.


\section{System Integration}

To properly assess its performance, a \gls{NOC} should be integrated into a full \gls{SOC} architecture and analyzed in context. We address this requirement in the following subsections.


\subsection{Compute Tile}

\begin{figure}[htbp]
    \centering
    \includegraphics[width=\columnwidth]{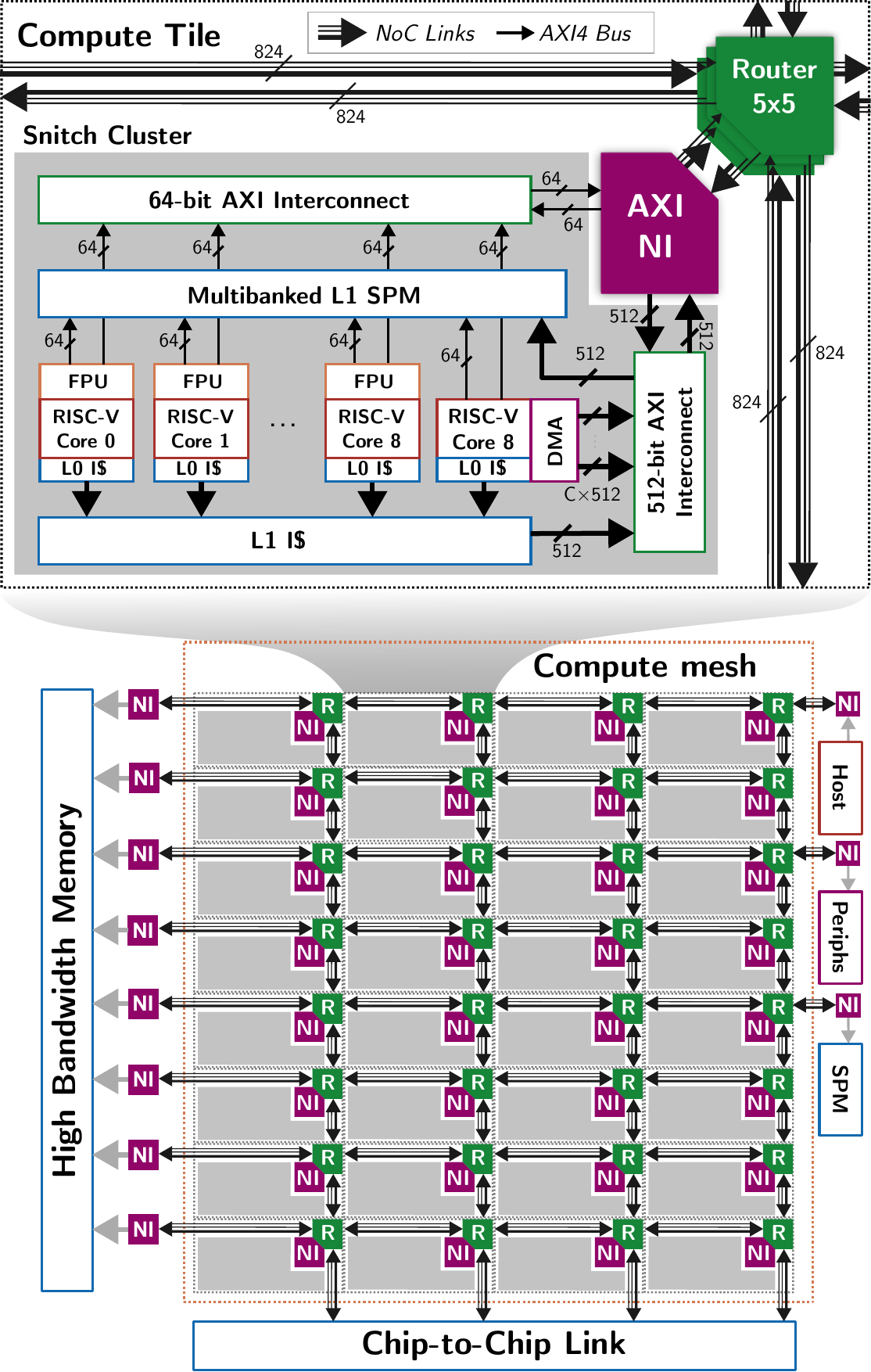}
    \caption{\textit{Top}: Compute tile consisting of a Snitch cluster with nine RISC-V compute cores, one of which is tightly-coupled to a \gls{DMA} engine capable of handling $C$ streams in parallel. An \gls{AXI}-\gls{NI} attached to the wide 512-bit and narrow 64-bit \gls{AXI} bus, and separate 5$\times$5 router for each physical link in the narrow-wide network. \textit{Bottom}: Compute mesh of 8$\times$4 compute tiles. The compute mesh connects to the \gls{HBM} on the left, to \gls{C2C}-link for off-chip communication on the bottom.}
    \label{fig:flooccamy}
\end{figure}


We integrate the \gls{AXI}-\gls{NI} and router into a Snitch cluster \cite{snitch}, creating a compute tile, as illustrated in \Cref{fig:flooccamy}. \removed{The router is configured for XY-Routing with 5$\times$5 ports: one local port for the cluster and one port for each cardinal direction.} The compute cluster comprises eight \added{lightweight, single-stage, in-order RISC-V cores, each tightly coupled with a pipelined, multi-precison, SIMD-capable \gls{FPU}.} \removed{RISC-V cores with integrated FPUs and a} \added{An additional} ninth RISC-V core \added{is} dedicated to controlling the \gls{DMA}. The \gls{SPM} and instruction cache, shared among all cores, are sized at \SI{128}{\kilo\byte} and \SI{8}{\kilo\byte}, respectively. The internal interconnect within the cluster includes a 512-bit wide \gls{AXI} bus used by the \gls{DMA} and L1 instruction cache to fetch large data blocks. Additionally, all RISC-V cores connect to a narrow 64-bit \gls{AXI} bus for single-word remote accesses. Both the narrow and wide \gls{AXI} buses are equipped with initiator and target ports, allowing the cluster’s internal \gls{SPM} to be accessed remotely by other clusters. The \gls{AXI}-\gls{NI} is connected to both the narrow and wide \gls{AXI} buses according to the mapping outlined in \Cref{tab:channels}. \added{The router is configured for XY-Routing with 5$\times$5 ports: one local port for the cluster and one port for each cardinal direction.}

\subsubsection{Ordering \& Multi-Channel DMA}
We enable out-of-order transactions in the cluster in multiple ways. For both the narrow and the wide \gls{AXI} interface, we configure the \gls{NI} with the newly proposed \textit{\gls{ROB}-less} version instead of our previous work~\cite{floonoc} for more cost-effective and performant end-to-end ordering. We prevent stalls that might occur from the ordering guarantee the following way: The narrow ports of each core in the Snitch cluster have unique IDs, which enables out-of-order responses from different cores by design. Only simultaneous accesses of the same core to different destinations might exhibit stalls. However, the compute cores mainly operate on the cluster-internal L1 \gls{SPM}, and narrow traffic out of the cluster is rare and limited to synchronization between clusters, which should not be affected by stalls.

The wide \gls{AXI} interface is the more critical component, which previously required a large \gls{ROB} buffer~\cite{floonoc} to ensure high performance. In this work, however, we extended the compute cluster with a multi-channel \gls{DMA}~\cite{benz2023highperformance} that allowed us to get rid of the \gls{ROB} inside the \gls{NI} and offload the ordering of different streams to the \gls{DMA} itself. The \gls{DMA} can be programmed through a single \textit{frontend} that controls multiple \textit{backends} of the \gls{DMA}, each capable of handling a separate stream. The \gls{DMA} has one wide 512-bit \gls{AXI} port for each \textit{backend}, which is connected to the wide \gls{AXI} \gls{XBAR}, meaning each stream coming from a \textit{backend} and entering the \gls{NI} has a unique \gls{TxnID} and does not imply any ordering with other streams.

\subsection{Compute Mesh}
The system can be scaled up by replicating the compute tile and arranging it as a homogeneous compute mesh of tiles, as shown in ~\Cref{fig:flooccamy}. The top-level connects the \gls{NOC} links of two neighboring tiles or ties them off if the tile is located at a boundary without any I/O components. Additional I/O components can be connected to the \gls{NOC} with an additional \gls{NI} that converts the \gls{NOC} protocol back to \gls{AXI} if required. For instance, the \gls{HBM} controller is attached to the left side of the compute mesh. We match the bandwidth to the \gls{HBM} by dimensioning the number of rows equal to the number of \gls{HBM} channels. For multi-chiplet scaling, a \gls{C2C} can be attached at the bottom of the compute mesh. Lastly, additional components such as a host processor, peripherals, and \gls{SPM} are attached to the right side of the compute mesh.

\begin{figure*}[ht]
  \centering
  \begin{subfigure}[b]{0.52\textwidth}
  \includegraphics[width=\columnwidth]{figures/pnr_annotated.pdf}
  \end{subfigure}
  \hfill
  \begin{subfigure}[b]{0.46\textwidth}
  \includegraphics[width=\columnwidth]{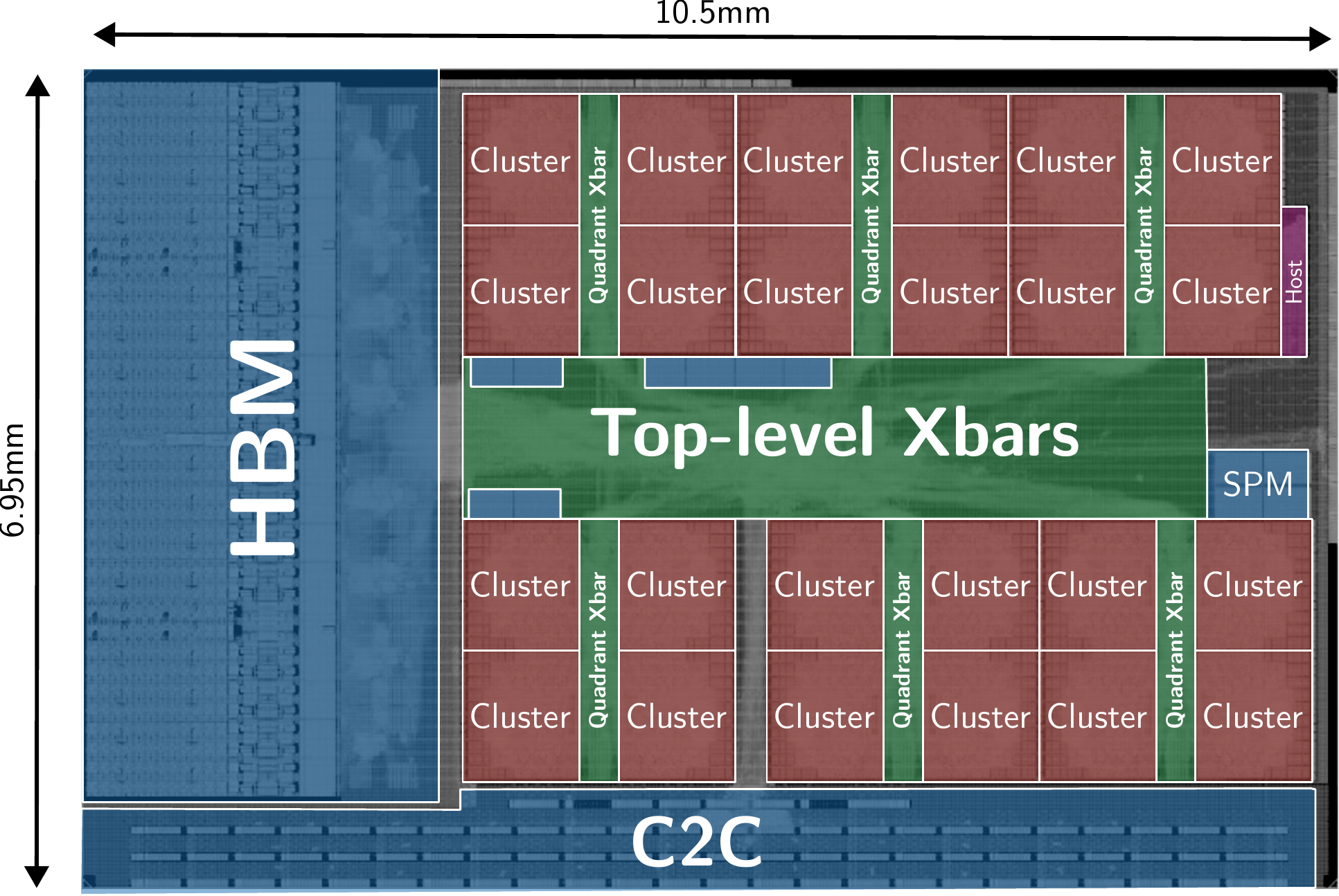}
  \end{subfigure}
  \caption{\added{Annotated placed and routed designs \textit{Left:} Macro of a single compute tile \textit{Middle: Compute Mesh with $8 \times 4$ } compute tiles \textit{Right:} Die shot of \textit{Occamy} including I/O components (\gls{HBM} and \gls{C2C}).}}
  \label{fig:pnr_annotated}
\end{figure*}

\section{Physical Implementation}

We performed the physical implementation of a compute mesh with $8 \times 4$ compute tiles. We used a hierarchical design flow with \textsc{Synopsys Fusion Compiler 2022.03} to synthesize, place, and route the compute mesh and tile in \textsc{GlobalFoundries'} \SI{12}{\nano\meter} FinFet technology.

\subsection{Compute Tile}

Most of the physical design effort was spent on the compute tile, which is replicated on the top-level to form a compute mesh. \removed{The floorplan of the tile was chosen to have an aspect ratio of 2:1 to maximize the horizontal bandwidth per coastline on the left side facing the HBM.} As observed in \cite{paulin2022soft}, the most challenging component of the Snitch cluster is the fully connected crossbar to the multi-banked L1 \gls{SPM} since it is routing-bound. In order to utilize the full routing resources available inside a tile, we aimed to separate the routing-intensive \gls{SPM} crossbar and the \gls{NOC} links. First, We placed the memory macros of the \gls{SPM} in a U-shape to the right of the floorplan to force the crossbar to the right side as well. Second, we placed the ports of the \gls{NOC} links on the left, respectively, on the top side of the edge to pull the router away from the routing-bound crossbar. The result is seen in the annotated floorplan of the tile in \Cref{fig:pnr_annotated}. The ports of the \gls{NOC} were deliberately placed on the upper layers of the metal stack so that the global wires of the \gls{NOC} links could easily be routed over the memory macros, which only occupy the bottom layers of the metal stack.

The input and output of the tile were constrained with \SI{75}{\percent} of the clock period to reflect the fact that the signal from the neighboring tile had already had to travel a large distance. In order to ease the timing of the \gls{NOC} links that need to bridge a distance $>$\SI{1}{\milli\meter}, we additionally configured the routers with output buffers. Further, we flattened the three routers directly into the compute tile instead of creating a macro. This has multiple benefits: 1) The router can absorb the under-utilized routing resources from the surrounding logic \cite{batten_noc_symbiosis} 2) The tool can spread out the input and output buffers over a larger area than a router macro, allowing it to meet the input and output timing more easily.

\subsection{Compute Mesh}

The physical design of the compute mesh mainly consisted of determining the aspect ratio of the \added{\textit{Compute Tiles} floorplan and their mesh arrangement.} \removed{floorplan and the placement of the \textit{Compute Tiles} in a mesh structure} The height of the \textit{Compute Mesh} was chosen to match the height of the \gls{HBM} controller used in \cite{occamy_vlsi_2024}. \added{Further, we aligned the number of rows in the mesh to the number of \gls{HBM} channels in order to maximize the \gls{HBM} bandwidth, which resulted in an asymmetric 2:1 aspect ratio for the tile floorplan.} \removed{Further}\added{Finally}, the compute tiles were placed with a small gap between the top-level \gls{NOC} connections. This gap is also used for routing the clock tree, which prevents a direct abutment of the tiles.

The entire \textit{Compute Mesh} is one synchronous clock domain, and the timing closure was done in two steps. First, all \textit{reg2reg} paths inside a \textit{Compute Tile} were closed, and second, all paths on the top-level. The \textit{in2reg} and \textit{reg2out} paths on the \textit{Compute Tile} level are not relevant for timing closure as long as the timing on all \textit{tile2tile} paths can be met at the top-level. The maximum frequency is then determined by the \gls{WNS} of \textit{Compute Tile} and \textit{Compute Mesh}.

\section{Results}

We first evaluate the performance of the \gls{NOC} in terms of latency and throughput and then discuss physical design results such as area, timing, and energy efficiency of the design. \textit{FlooNoC} is implemented as SystemVerilog RTL, and performance was simulated with \textsc{QuestaSim 2023.4}. Performance numbers were extracted from cycle-accurate traces of accesses issued by the cores and \glspl{DMA}. Further, we used \textit{DRAMSys}~\cite{jung2015dramsys} with a configuration of eight Micron MT54A16G808A00AC-36 \gls{HBM}2E channels to accurately model the latency and throughput of \gls{HBM} accesses. The \gls{HBM}2E model we used has a peak bandwidth of \SI{57.6}{\giga\byte\per\second} per channel.

The area and timing numbers were extracted from the netlist of the placed and routed design in \textsc{GlobalFoundries'} \SI{12}{\nano\meter} FinFet technology with \gls{STA} of \textsc{Synopsys Fusion Compiler 2022.03}. Further, the same netlist was used to perform post-layout power simulations using \textsc{Synopsys PrimeTime 2022.3} in typical conditions (TT, \SI{0.8}{\volt}, \SI{25}{\celsius}).

\begin{figure}[ht]
    \centering
    \includegraphics[width=\columnwidth]{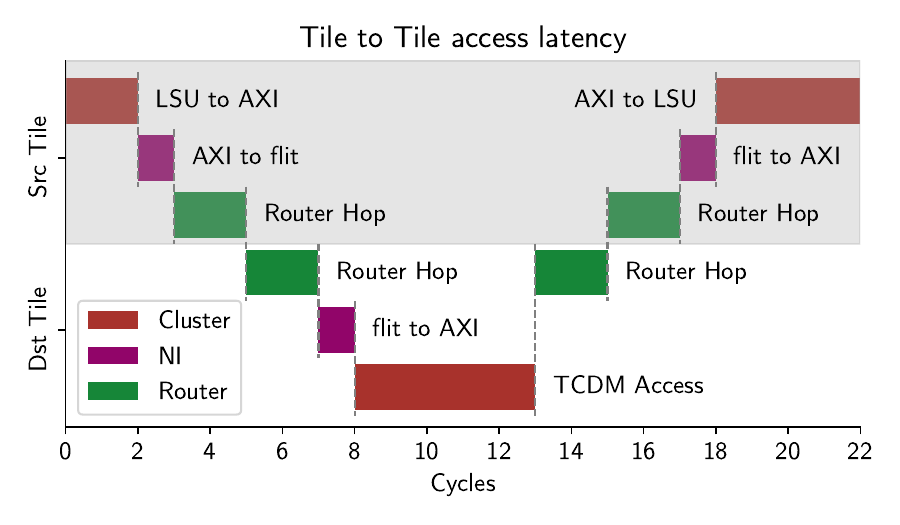}
    \caption{Latency breakdown of a Tile to Tile access of a narrow transfer.}
    \label{fig:lat_gantt}
\end{figure}

\begin{figure}[ht]
    \centering
    \includegraphics[width=\columnwidth]{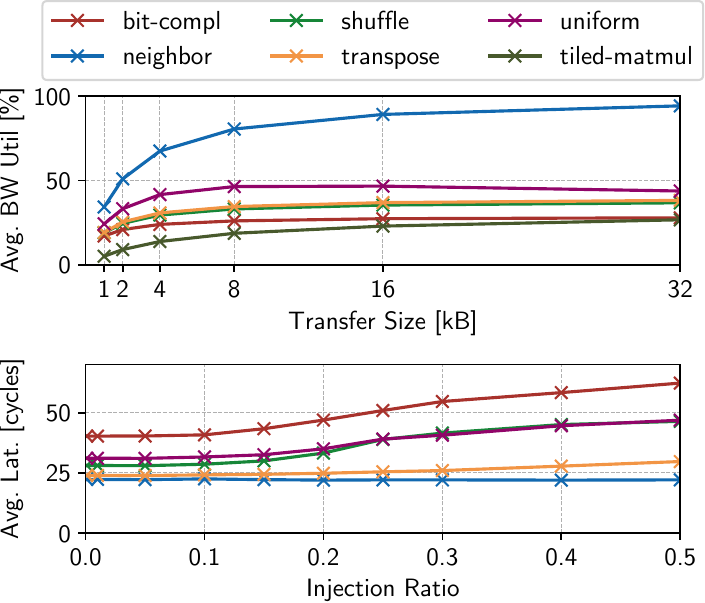}
    \caption[Caption without FN]{\added{
            Compute mesh performance for different traffic patterns  \textit{Top:} Average bandwidth utilization of wide accesses \textit{Bottom:} Average latency of narrow accesses}\footnotemark}
    \label{fig:traffic_patterns}
\end{figure}

\subsection{Latency}

We measured the latency experienced by cores accessing the L1 \gls{SPM} of another tile over the narrow \gls{NOC} links. For instance, this type of traffic occurs during the synchronization of multiple clusters, which is much more latency-sensitive. A breakdown of the Tile-to-Tile access latency is shown in \Cref{fig:lat_gantt}.

Accessing a neighboring tile \removed{during \textit{zero-load}} results in a latency of 22 cycles, which includes the entire roundtrip of sending a read request and receiving the read data. Only eight of the total cycles are contributed from the \gls{NOC} routers, where each hop has a cost of a two-cycle latency. Additionally, the \glspl{NI} add three cycles of latency for translating an \gls{AXI} request or response to a flit and vice-versa. The remaining latency comes from the cluster-internal interconnect as well as memory access latency. Each additional hop to access a non-neighboring tile costs four cycles \removed{. In the worst case, communication between two tiles can cost up to 58 cycles in an $8 \times 4$ mesh}\added{, resulting in a latency of 58 cycles for tiles located at opposite corners of the $8 \times 4$ mesh.} \removed{The latency significantly increases when \textit{full-load} is considered and can reach up to 321 cycle latency in the worst case when every core (a total of 279 cores) is accessing the same tile through the {\gls{NOC}}. Note that this very unlikely scenario can be easily avoided in software. For instance, synchronization of all cores can be done hierarchically with intra-cluster followed by inter-cluster synchronization, which decreases the number of simultaneous accesses by $9\times$.}

\added{For more complex traffic patterns that involve the entire compute mesh, as shown in \Cref{fig:traffic_patterns}, latency moderately increases for patterns that cause congestion. In contrast, for localized traffic without contention, such as \textit{neighbor} traffic, the \gls{NOC} is capable of injecting one packet per cycle without any latency degradation.}

\subsection{Bandwidth}

\footnotetext{\added{The latency of the \textit{tiled-matmul} pattern is omitted, since it involves high-latency \gls{HBM} accesses which are not representative of the \gls{NOC} latency.}}

Our \gls{NOC} solution aims to provide a sustained high bandwidth data flow required by many applications today~\cite{squeezeLLMmemorybound}. The cluster-internal \gls{DMA} can handle large bursts of data while providing tolerance to latency due to the ability to issue multiple outstanding transactions at once. The 512-bit wide links in the \gls{NOC} used by the \glspl{DMA} achieve a peak bandwidth of \SI{645}{\giga\bitpersecond} (\SI{1.29}{\tera\bitpersecond} duplex), operating at a frequency of \SI{1.26}{\giga\hertz}. \removed{This allows for considerable bandwidth across the network, specifically tailored to handle the demands of high-volume data traffic directed toward memory controllers and I/O interfaces such as HBM controllers and C2C links. In the practical application within an $8 \times 4$ compute mesh, the NoC demonstrates an effective delivery of massive bandwidth at the boundary.}

\added{We also evaluate the bandwidth utilization achieved for multiple popular synthetic traffic patterns on the $8 \times 4$ compute mesh, which are shown in \Cref{fig:traffic_patterns}. Bandwidth utilization mainly depends on the transfer size and the level of network congestion. For instance, \textit{neighbor} traffic, which experiences zero contention, achieves nearly peak bandwidth utilization for \SI{32}{\kilo\byte} transfers. In contrast, highly congested traffic patterns, such as \textit{bit-complement}, reach only 28\% utilization. Additionally, we simulated a \textit{tiled-matmul} pattern, which emulates the access patterns of tiled matrix multiplication, which is representative of many \gls{AI} workloads. This traffic is dominated by read traffic to \gls{HBM} and fewer writes to \gls{HBM} to store back the resulting matrix tile. Note that this traffic pattern is limited by the \gls{HBM} bandwidth at the boundary.}

\removed{We also evaluate the bandwidth utilization achieved at the boundary when accessing HBM channels. We simulate the maximum bandwidth achieved where each tile accesses the {\gls{HBM}} channel of its row in the mesh. Additionally, we simulate two different conditions: \textit{zero-load}, where each tile is the only accessor of the {\gls{HBM}} channel, and \textit{full-load}, where all tiles in a row simultaneously access an {\gls{HBM}} channel, representing a maximum contention scenario on each {\gls{HBM}} channel.}
\removed{The results are shown in Figure 11. Under the zero-load condition, the bandwidth utilization of an {\gls{HBM}}2E channel ({\SI{57.6}{\giga\byte\per\second}}) is almost maximized, with most channels achieving near-optimal utilization levels of 97\%. The observed decrease to 91\% in one tile is due to increased contention with instruction fetches on the first {\gls{HBM}} channel where the test binary is stored.}
\removed{These instruction fetches are counted towards the bandwidth utilization. In the full-load condition, each tile achieves up to 28\% {\gls{HBM}} bandwidth utilization and a full combined HBM bandwidth utilization in each row of the mesh. It is also noteworthy that the bandwidth distribution amongst the column of the tiles is fair, even if requests from the right tiles face more contention than tiles closer to the {\gls{HBM}} channels, which has been reported to be a problem when using standard RR arbiters in the routers \mbox{\cite{tvlsi_wrr_arb}}.}

\begin{figure*}[htbp]
    \centering
    \begin{subfigure}[b]{0.48\textwidth}
      \centering
      \includegraphics[width=\columnwidth]{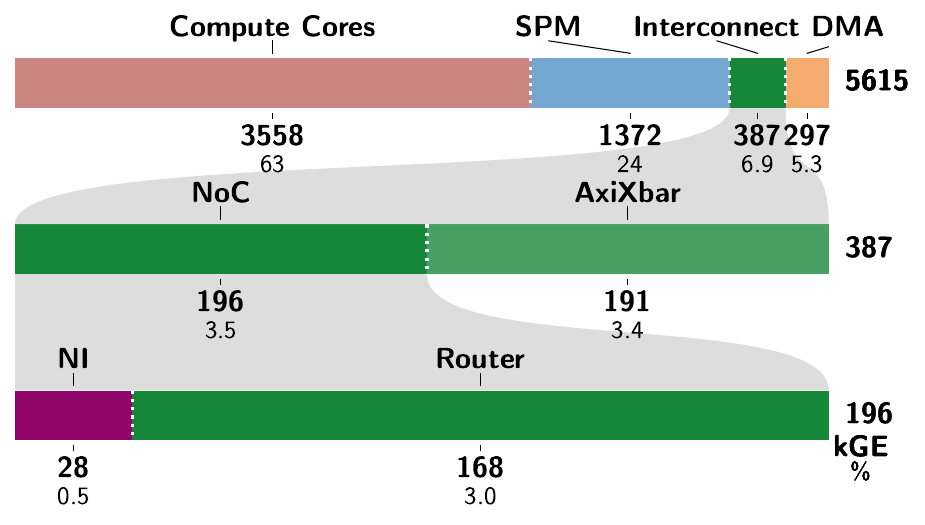}
      \caption{Area breakdown}
      \label{fig:area}
    \end{subfigure}
    \hfill
    \begin{subfigure}[b]{0.48\textwidth}
      \centering
      \includegraphics[width=\columnwidth]{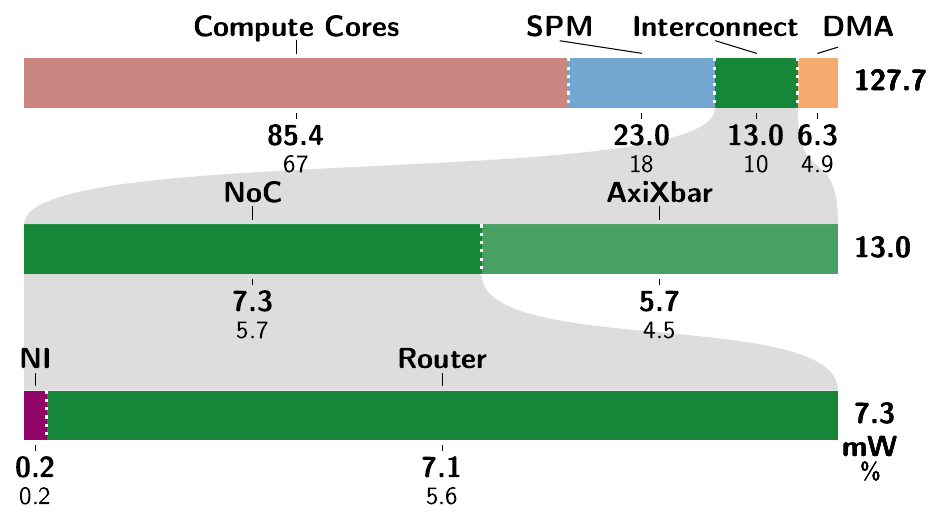}
      \caption{Power breakdown}
      \label{fig:power}
    \end{subfigure}
    \label{fig:breakdowns}
    \caption{post-layout area and power breakdowns of a single tile (a) Area breakdown of a single tile with integrated \gls{NOC} components (b) Power breakdown of a tile during a \SI{4}{\kilo\byte} \gls{DMA} transfer from tile to neighboring tile. All cores expect the \gls{DMA} core to remain idle during the transfer.}
\end{figure*}

\subsection{Area \& Timing}

The area impact of the \gls{NOC} components is small, as shown in the area breakdown in \Cref{fig:area}. The largest contributors to the total area of the tile are the compute cores, with the \glspl{FPU}, followed by the L1 \gls{SPM}. The entire interconnect, which includes the \gls{NOC} and the cluster-internal wide 512-bit \gls{AXI} \gls{XBAR}, only occupies 6.9\% of the area, which is almost evenly split between \gls{NOC} and \gls{XBAR}. The \gls{NOC} complexity is largely dominated by the router, where 53\% of the area is taken up by the \gls{SCM} input and output buffers, while the rest is needed for the router switch.

\begin{figure}[htbp]
    \centering
    \includegraphics[width=\columnwidth]{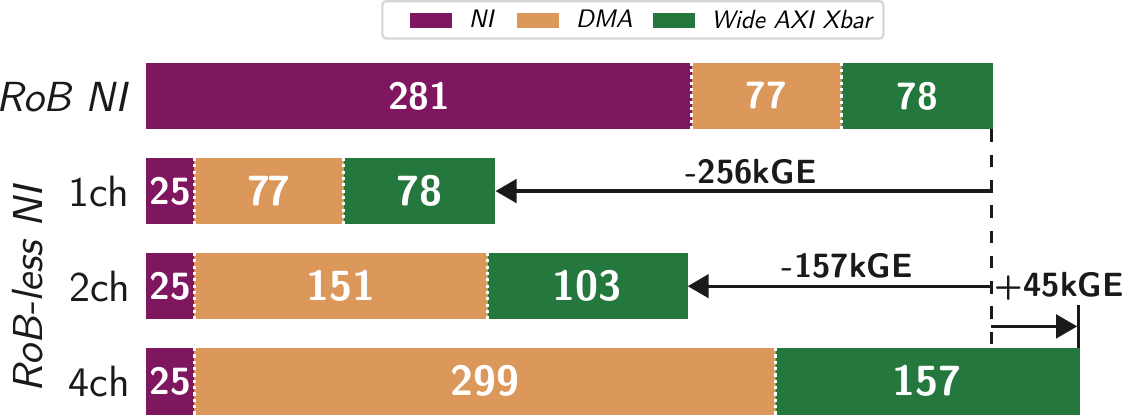}
    \caption{Area breakdown in \SI{}{\kilo\gate} of components in a tile affected by end-to-end ordering. Comparison between \textit{\gls{ROB}} and \textit{\gls{ROB}-less} configurations with 1-4 \gls{DMA} channels are shown. The \gls{ROB} has a size of \SI{8}{\kilo\byte} implemented as \glspl{SRAM}.}
    \label{fig:area_ni_dma_rob}
\end{figure}

The \gls{NI} without a \gls{ROB} is almost negligible in complexity with\added{in} a compute tile, taking up only \SI{25}{\kilo\gate}. This is possible thanks to the end-to-end optimization of \gls{AXI} ordering in the cluster's \gls{DMA}. \added{Compared to the previous \gls{ROB}-based \gls{NI} \cite{floonoc}, this results in a \SI{256}{\kilo\gate} area reduction, representing a 91\% decrease with a single-channel \gls{DMA} configuration, as shown in \Cref{fig:area_ni_dma_rob}.}
\added{However, enabling out-of-order transactions requires a multi-channel \gls{DMA}, which increases the complexity of both the \gls{DMA} and the wide \gls{AXI} \gls{XBAR} due to the additional ports.}
\removed{However, enabling multi-channel DMA comes at a cost, which is shown in Figure 10. the DMA complexity increases significantly for a 4-channel configuration, as it needs to handle multiple streams in parallel. Furthermore, each channel has its own AXI port to the cluster internal wide AXI Xbar, which causes another 50\% increase in the size of the Xbar.}
Consequently, the area reduction in the \gls{NI} is compensated in the \gls{DMA} and \gls{XBAR}, but with the benefit of having multiple parallel streams that do not interfere with each other and without performance degradation resulting from limited \gls{ROB} capacity.

The critical path of the compute mesh is inside the Snitch cores of the compute cluster and is not affected by the \gls{NOC}. The timing closes at \SI{1.26}{\giga\hertz} in typical conditions (TT, \SI{0.8}{\volt}, \SI{25}{\celsius}), which corresponds to a delay of 67 \gls{FO4}.


\begin{figure*}[ht!]
    \centering
    \begin{subfigure}[b]{0.48\textwidth}
        \centering
        \includegraphics[width=\textwidth]{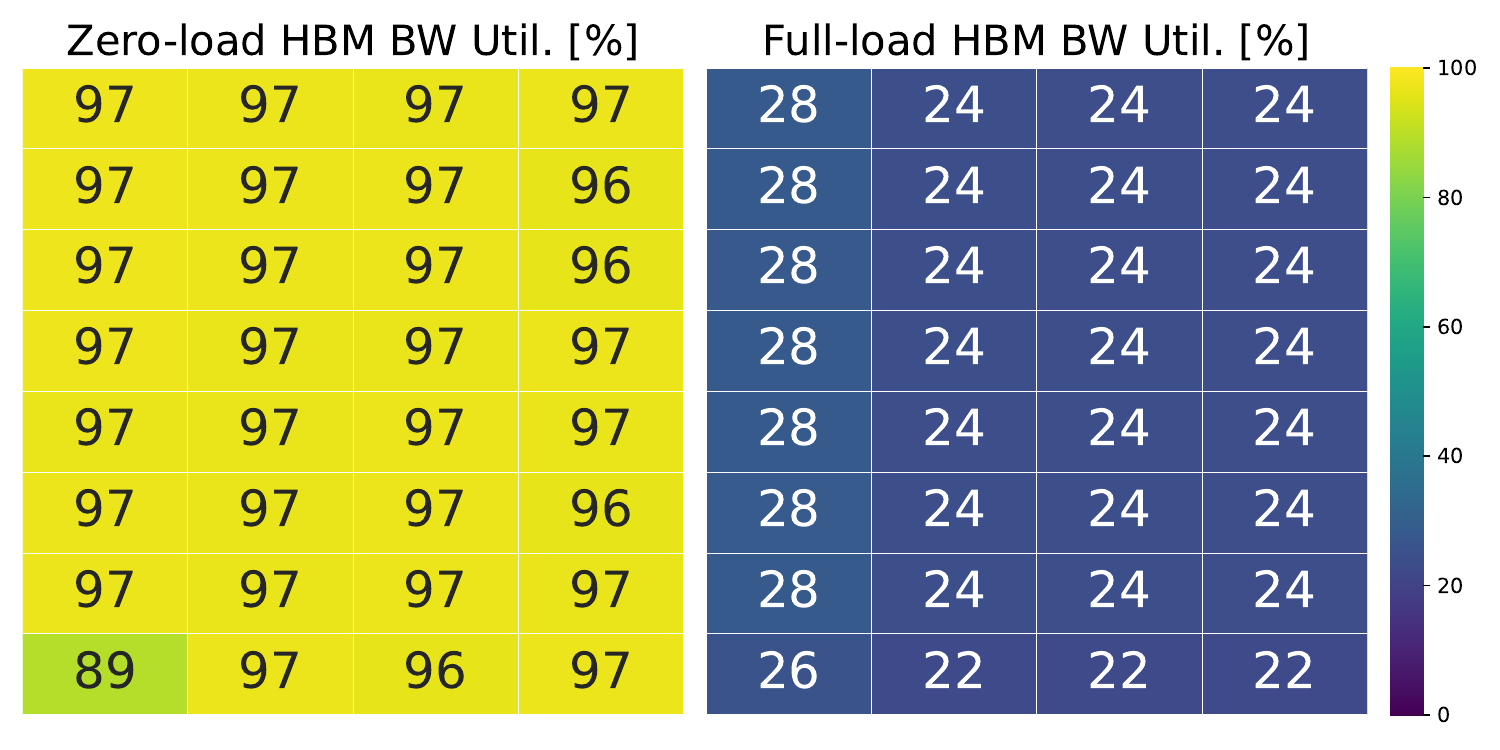}
    \caption{Wide link utilization relative to the maximum \gls{HBM} bandwidth of a channel in \textit{FlooNoC} when each tile accesses one \gls{HBM} channel (\texttt{tile\_coord\_y}).}
        \label{fig:bw_flooccamy}
    \end{subfigure}
    \hfill
    \begin{subfigure}[b]{0.48\textwidth}
        \centering
        \includegraphics[width=\textwidth]{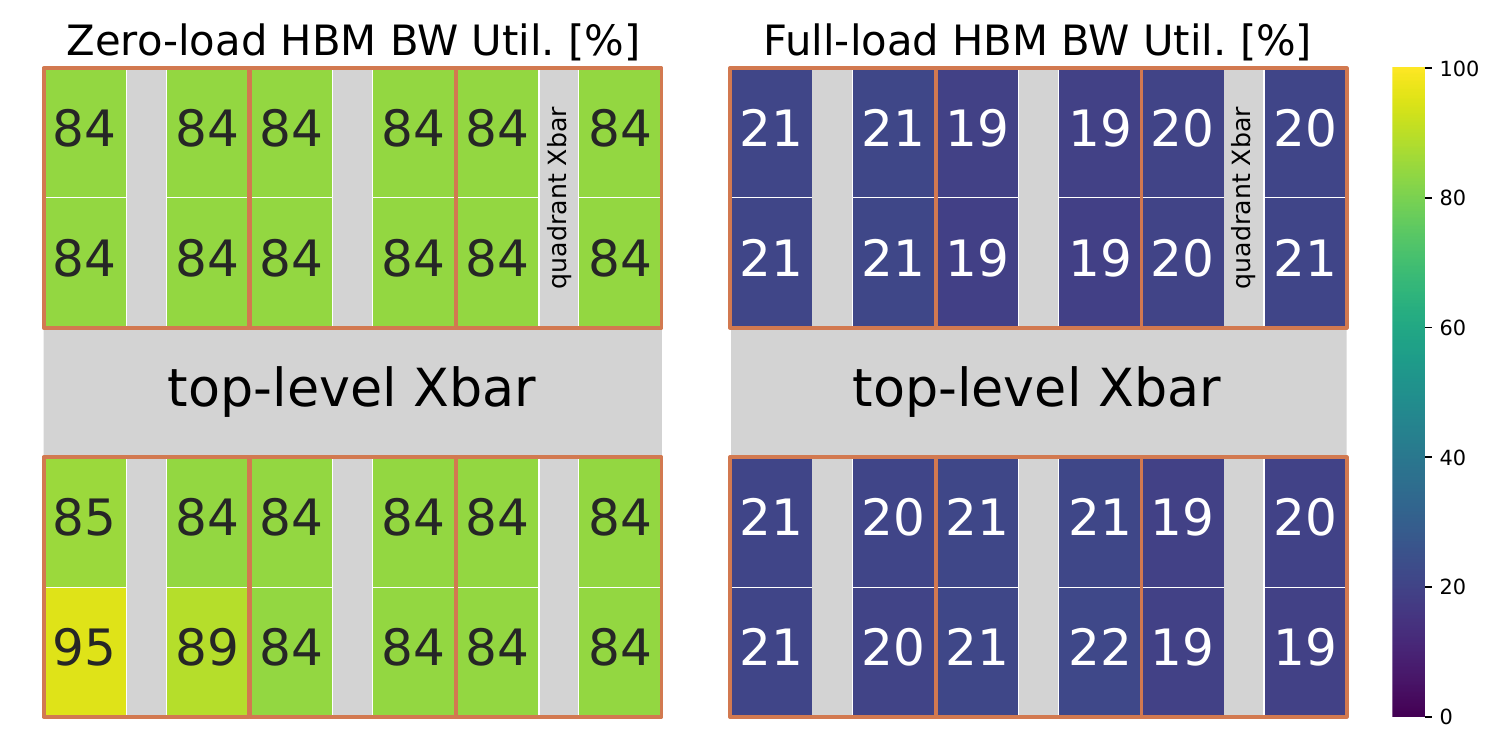}
        \caption{Utilization of the wide \gls{AXI}-bus relative to the maximum \gls{HBM} bandwidth of a channel in \textit{Occamy} when each cluster accesses one \gls{HBM} channel (\texttt{cluster\_id \% \#HBM\_CHs}).}
        \label{fig:bw_occamy}
    \end{subfigure}

    \begin{subfigure}[b]{0.48\textwidth}
        \centering
        \includegraphics[width=\textwidth]{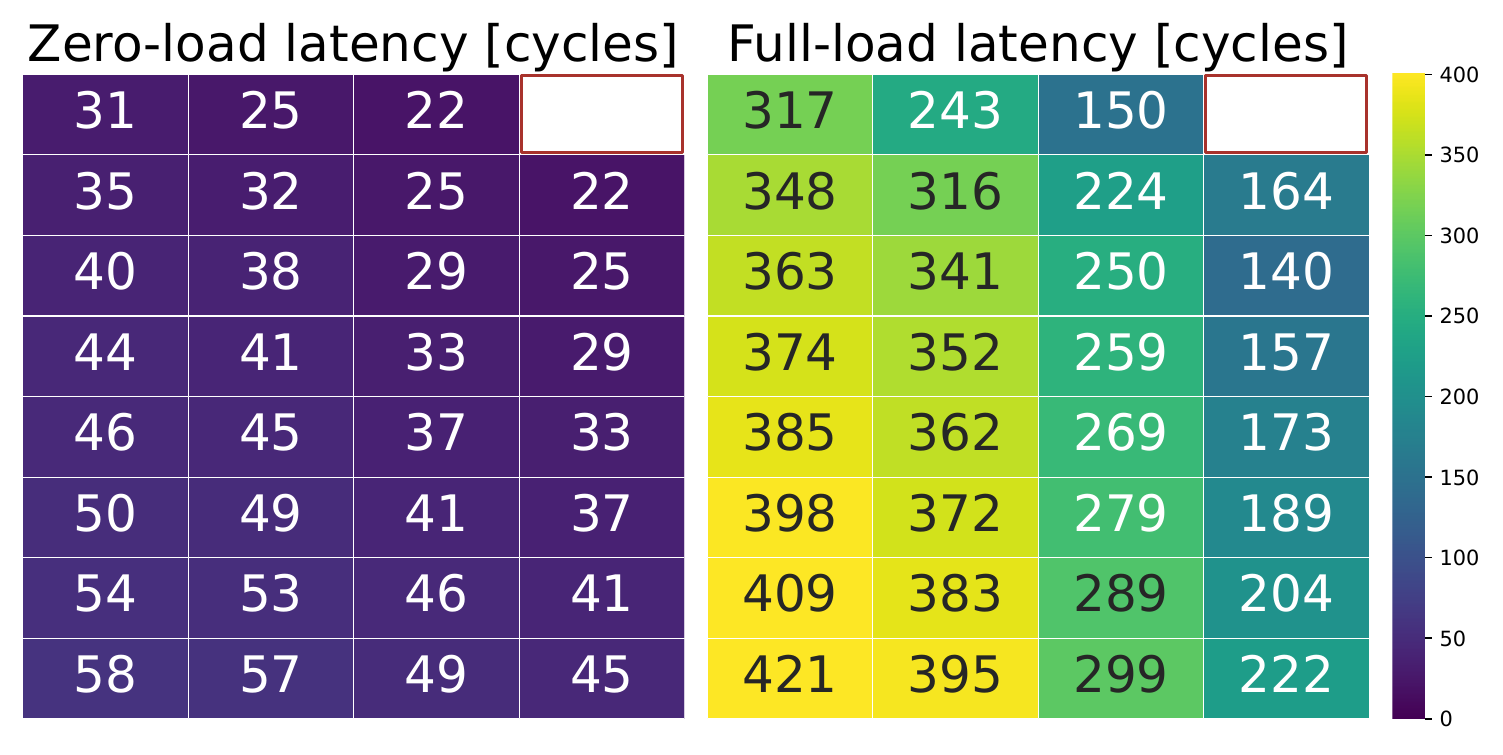}
        \caption{Latency of narrow reads to top right tile in \textit{FlooNoC}}
        \label{fig:lat_flooccamy}
    \end{subfigure}
    \hfill
    \begin{subfigure}[b]{0.48\textwidth}
        \centering
        \includegraphics[width=\textwidth]{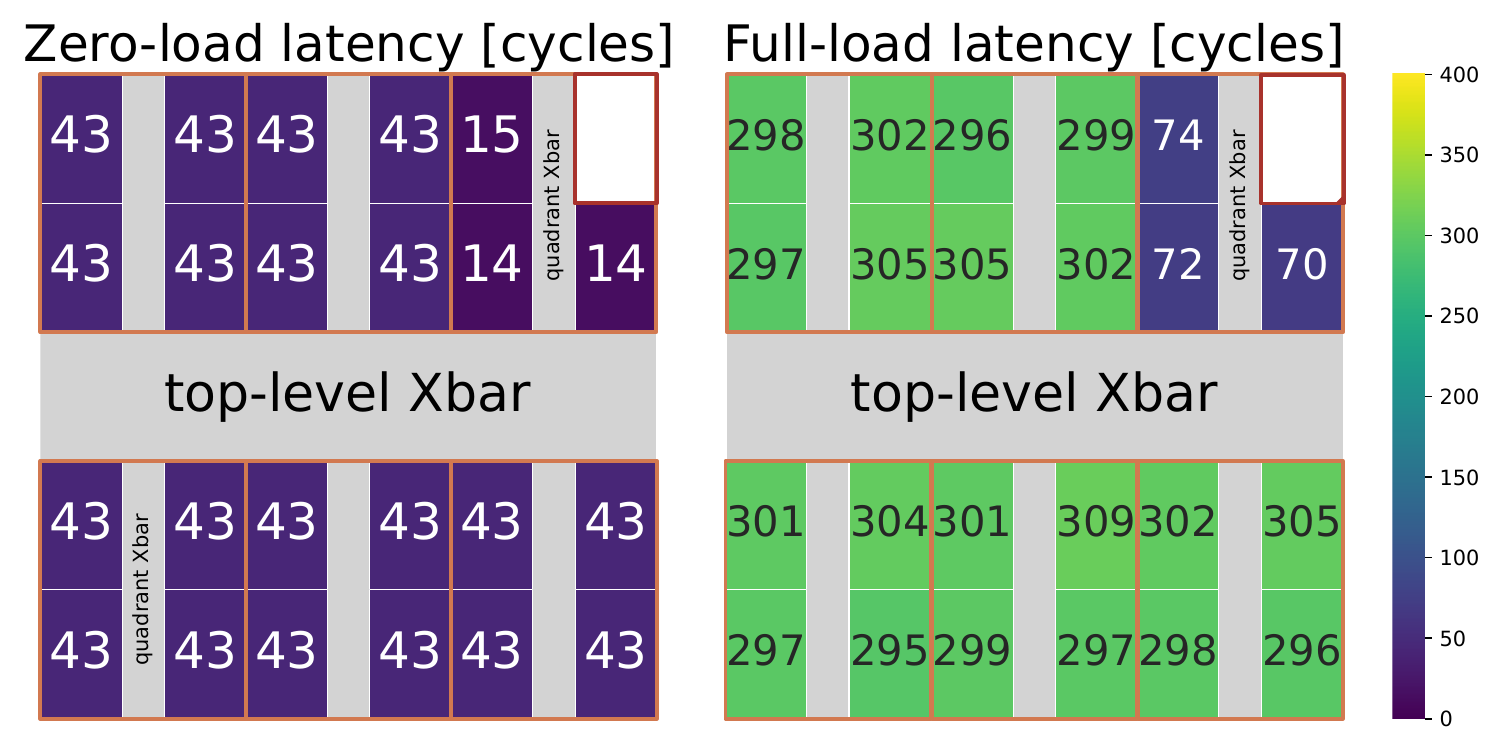}
        \caption{Latency of narrow reads to top right cluster in \textit{Occamy}}
        \label{fig:lat_occamy}
    \end{subfigure}
    \caption{Latency of narrow accesses and bandwidth utilization of the wide links in the $8\times4$ compute mesh and $6\times4$ \textit{Occamy} organized by their physical location. During \textit{zero-load} only one core or \gls{DMA} issues requests and \textit{full-load} all cores or \glspl{DMA} issue request simultaneously. In the bandwidth simulations, each \gls{DMA} issues \textsc{32Txns} of size \textsc{4KB}.}
    \label{fig:bw_lat_heatmaps}
\end{figure*}

\subsection{Energy \& Power}

In our analysis, we simulated the power consumption involved in transferring \SI{4}{\kilo\byte} of data from one tile to a neighboring tile over the \gls{NOC}, as detailed in~\Cref{fig:power}. Although all eight compute cores remained idle during this transfer, they were the primary contributors to power consumption. This substantial usage is primarily due to the clock tree, which accounts for nearly 50\% of the idle power usage, and leakage power, which constitutes another 10\%. As a result, only about 15\% of the total power of \SI{127.7}{\milli\watt} is consumed by the components actively involved in the transfer (\gls{DMA}, wide \gls{AXI} \gls{XBAR} \& \gls{NOC}). Consequently, the \gls{NOC} is not the limiting factor in terms of power consumption on a system level. Furthermore, the routers only consume \SI{596}{\pico\joule} during the transfer, resulting in an energy efficiency of \SI{0.15}{\pico\joule\per\byte\per\hop}.

\section{System-Level Comparison}

We also performed a system-level comparison with \textit{Occamy}, which features the same compute clusters but connects them with a hierarchical \gls{AXI} interconnect instead of a mesh, as is shown in the annotated die shot in \Cref{fig:pnr_annotated}. On the lowest level of the hierarchy, \textit{Occamy} groups together four clusters with an \gls{AXI} \gls{XBAR} for both the 64-bit narrow and the 512-bit wide interconnect to form a group. Further, six of those \textit{groups} are connected with a top-level \gls{XBAR} that also connects the system with the \gls{HBM} and \gls{C2C} link. We measured the performance of the \textit{Occamy} system with the same setup, i.e., extracting the performance from cycle-accurate traces from RTL simulations and \gls{HBM} modeling with \textit{DRAMSys}. The area and timing numbers were generated from the area reports, respectively \gls{STA}, from the database of the final design that was taped out.

\subsection{Bandwidth}
First, we compare the bandwidth of \textit{Occamy} against \textit{FlooNoC}-based systems. On a basic level, both designs rely on the wide and duplex 512-bit \gls{AXI} bus for high-bandwidth traffic. However, the wide link in \textit{FlooNoC} is shared amongst read and write data in one direction, as explained in \Cref{subsec:link_flits}, as opposed to a standard \gls{AXI} bus, which defines separate R and W channels in each direction. In theory, the link sharing in \textit{FlooNoC} would halve the bandwidth for concurrent read and write access from and to a tile. However, the L1 \gls{SPM} has a bandwidth of only \SI{512}{\bit\per\cycle}, limiting the full utilization of R and W channels. Furthermore, the bandwidth achieved on a wide link (\SI{645}{\giga\bitpersecond}) is slightly higher than in \textit{Occamy} (\SI{584}{\giga\bitpersecond}) due to the slightly higher maximum frequency.

We also compared the bandwidth utilization of the \gls{HBM} channels that can be achieved in both systems, which is shown in \Cref{fig:bw_flooccamy} and \Cref{fig:bw_occamy}. In both the \textit{zero-load} and \textit{full-load} scenarios, \textit{FlooNoC} achieves higher utilization. The reasons for the difference in bandwidth utilization are manyfold. First, the top-level \gls{XBAR} in \textit{Occamy} had to be split up into a hierarchy of \glspl{XBAR} to make it even possible to implement physically, which increases the depth of the memory hierarchy and, consequently, the number of hops and latency. Second, the physical distances between \glspl{XBAR} in a hierarchical interconnect are much larger compared to a mesh that only connects neighboring tiles. Multiple spill registers must be inserted to bridge this distance, which again increases latency. Even though \gls{AXI} supports multiple outstanding transactions to tolerate latency, the \textit{zero-load} results show that increased latency can still impact the bandwidth. In \textit{FlooNoC}, the latency to the nearest \gls{HBM} channel is much lower, resulting in higher bandwidth overall.

The difference in \gls{HBM} bandwidth utilization in the \textit{full-load} is even more significant. Even though \textit{FlooNoC} has more clusters than \textit{Occamy}, a cluster in \textit{FlooNoC} gets a larger share of the \gls{HBM} bandwidth. The most obvious reason is that \textit{FlooNoC} has one link to each \gls{HBM} channel, while \textit{Occamy} only has six (one from each cluster). Moreover, the \gls{AXI} \glspl{XBAR} in \textit{Occamy} do not support as many outstanding transactions as would be needed in the \textit{full-load} case. In a hierarchical interconnect, \gls{AXI} \glspl{XBAR} need to be configured to support progressively more outstanding transactions with each additional level, which becomes too expensive at some point. These issues limit the \textit{full-load} \gls{HBM} bandwidth utilization in \textit{Occamy} to around 60\% compared to near-maximum utilization in \textit{FlooNoC}. \added{It is also noteworthy that the bandwidth distribution amongst the column of the tiles is fair, even if requests from the right tiles face more contention than tiles closer to the \gls{HBM} channels, which has been reported to be a problem when using standard \gls{RR} arbiters in the routers \cite{tvlsi_wrr_arb}.}

\subsection{Latency}

Regarding latency, the hierarchical interconnect of \textit{Occamy} has its merits in some cases. As is shown in \Cref{fig:lat_flooccamy} and \Cref{fig:lat_occamy}, the cluster-to-cluster access latency is slightly lower in \textit{Occamy} when the access is happening inside a \textit{group} since it only consists of a single hop of an \gls{XBAR} and also omits the latency induced by the \glspl{NI}. On the other hand, access to a cluster in another \textit{group} is much more costly, with 43 cycles of latency. The group-to-group latency in \textit{Occamy} also does not depend on the location of the other \textit{group}, which is more relevant in a mesh-based system where the latency depends on the number of hops throughout the network. This is visible in both the \textit{zero-load} and \textit{full-load} scenarios. While \textit{FlooNoC} has a larger maximum latency in both scenarios, it represents the worst-case latency and can more easily be mitigated with physical awareness of the data. For instance, placing a synchronization barrier in the middle of the mesh will have an immediate positive effect on latency, while the latency in \textit{Occamy} will remain constant and independent of the choice of cluster or \textit{group}.

\subsection{Area \& Timing}

\textit{FlooNoC} demonstrates a notable improvement in area efficiency over \textit{Occamy}, as detailed in \Cref{tab:occamy_comparison}. The die shot in \Cref{fig:pnr_annotated} clearly shows that the top-level \gls{XBAR} in \textit{Occamy} occupies almost 40\% of the compute domain\footnote{The compute domain only includes the cluster and \gls{AXI} interconnects without \gls{HBM} Ctrl./PHY and \gls{C2C}-link}. In contrast, an $8 \times 3$ compute mesh using FlooNoC achieves an 85\% reduction in top-level area and 30\% overall. Although integrating \gls{NOC} components increases the complexity of a tile slightly by 8\%, this is more than offset by the significant area savings at the top level. Moreover, an $8 \times 4$ compute mesh fits within the same floorplan as the compute domain in \textit{Occamy}, which only features 24 clusters.

    
The integration of \textit{FlooNoC} into the system does not adversely affect the overall timing in typical conditions, as the critical path in both systems resides in the compute cluster. However, \textit{FlooNoC} achieves a slightly higher frequency of \SI{1.26}{\giga\hertz} compared to \SI{1.14}{\giga\hertz} in \textit{Occamy}, which can be attributed to the slightly larger floorplan of the compute tile compared to the cluster floorplan in \textit{Occamy}. This results in a performance improvement of 10\% in terms of \SI{}{\giga\flops\double} for a 24-cluster system. Furthermore, \textit{FlooNoC} achieves a significantly higher compute density of \SI{16.4}{\giga\flops\double\per\square\milli\meter}, 58\% higher than \textit{Occamy}. This allows the integration of an additional eight clusters, which still fit into the same floorplan as \textit{Occamy}.


\begin{table}
    \centering
    \caption{Comparison with \gls{AXI}-Xbar based Occamy and compute mesh based on \textit{FlooNoC}}
    \begin{tabular}{@{}lccc@{}}
    \toprule
    & & \multicolumn{2}{c}{FlooNoC} \\
    \cmidrule{3-4}
    \textbf{Metric} & Occamy$^a$~\cite{occamy_vlsi_2024} & 8 $\times$ 3 Mesh & 8 $\times$ 4 Mesh \\
    \midrule
    \#Clusters & 24 & 24 (\textbf{+0\%}) & 32 (\textbf{+33\%}) \\
    Peak \SI{}{\giga\flops\double}$^b$ & 438 & 484 (\textbf{+10\%}) & 645 (\textbf{+47\%})\\
    Peak BW to \gls{HBM} [\SI{}{\tera\bitpersecond}] & 7.0 & 8.2 (\textbf{+17\%}) & 8.2 (\textbf{+17\%}) \\
    \midrule
    SS freq. [\SI{}{\giga\hertz}] & 0.88 & 0.90 (\textbf{+2\%}) & 0.85 (\textbf{-3\%})\\
    TT freq. [\SI{}{\giga\hertz}] & 1.14 & 1.26 (\textbf{+11\%}) & 1.26 (\textbf{+11\%}) \\
    \midrule
    Die Area [\SI{}{\square\milli\meter}] & 42.1 & 29.5 (\textbf{-30\%}) & 39.3 (\textbf{-7\%}) \\
    Tile/Cluster Area [\SI{}{\square\milli\meter}] & 25.1 & 27.0 (\textbf{+8\%}) & 36.0 (\textbf{+43\%})\\
    Top-level Area [\SI{}{\square\milli\meter}] & 16.7 & 2.5 (\textbf{-85\%}) & 3.3 (\textbf{-80\%})\\
    \midrule
    Compute density & \multirow{2}{*}{10.4} & \multirow{2}{*}{16.4 (\textbf{+58\%})} & \multirow{2}{*}{16.4 (\textbf{+58\%})}\\
    {[\SI{}{\giga\flops\double\per\square\milli\meter}]} & & & \\
    \bottomrule
    \end{tabular}
    \\
    $^a$considering only the compute domain of a single chiplet w/o \gls{HBM}, \gls{C2C} \\
    $^b$ TT frequency
    \label{tab:occamy_comparison}
\end{table}

\section{Comparison with SoA}

We also compare our physical implementation of the $8 \times 4$ compute mesh with the \gls{SOA} in \Cref{tab:soa_comparison}. Most of the previous work uses 2D-mesh topology, mainly due to the advantages during physical implementation. \textit{Piton} \cite{mckeown2018power}, for instance, employs a $5 \times 5$ mesh of tiles of cores with routers for three physical channels. While their \gls{NOC} is very inexpensive in terms of area utilization, the narrow 64-bit links cannot match the bandwidth provided by the wide links of \textit{FlooNoC}. They also report silicon measurements for energy efficiency in a \SI{32}{\nano\meter} technology, which are $3\times$ higher than our power simulations. More recent work has optimized the \textit{Piton} \gls{NOC} for \gls{HPC}~\cite{openpiton_hpc}, with wider links up to \SI{512}{\bit}, but is missing a physical implementation. The \textit{Celerity} chip \cite{celerity_19} achieves a very high aggregate \gls{NOC} bandwidth thanks to its more fine-grained tiling approach. However, they end up with a much lower tile-to-tile bandwidth that is ultimately limited by the very narrow link size of \SI{32}{\bit}. Another, more exploratory, work on physical implementation \cite{batten_pd_nocs20} used larger link sizes up to \SI{256}{\bit} to achieve a higher tile-to-tile bandwidth up to \SI{256}{\giga\bitpersecond}. However, the tile in which they integrated the \gls{NOC} is very small, resulting in a very high area overhead of more than a third of the tile area. Lastly, the \textit{ESP}-\gls{NOC} uses a slightly different approach, using six physical channels that achieve a respectable tile-to-tile bandwidth of \SI{310}{\giga\bitpersecond}. However, the \gls{NOC} contributes 23\% to the total \gls{SOC}'s power consumption of \SI{501}{\milli\watt}, and the energy efficiency cannot match \textit{FlooNoC}'s.

\newcommand*\rot[1]{\hbox to1em{\hss\rotatebox[origin=br]{-60}{#1}}}
\newcommand*\haeggli{\Circled[inner color=white, outer color=white, fill color=PULPGreen]{\cmark}}
\newcommand*\chruezli{\Circled[inner color=white, outer color=white, fill color=PULPRed]{\xmark}}
\newcommand*\welleli{\Circled[inner color=white, outer color=white, fill color=PULPOrange, inner ysep=6pt]{$\approx$}}

\begin{table*}[ht]
    \centering
    \caption{Comparison of State-of-the-Art systems with \glspl{NOC}}
    \begin{tabular}{@{}lcccccc@{}}
    \toprule
    \textbf{Work} & \textbf{Piton\cite{mckeown2018power}}& \textbf{Celerity\cite{celerity_19}} & \textbf{Ou et al.\cite{batten_pd_nocs20}} & \textbf{ESP\cite{esp_isscc24}} & \textbf{Prev. work\cite{floonoc}} & \textbf{This work} \\
    \midrule
    Technology & 32nm SOI & 16nm FinFET & 14nm & 12nm FinFET & 12nm FinFET & 12nm FinFET\\
    Voltage & 1.0V & \SI{0.98}{\volt} & - & \SI{0.8}{\volt} & \SI{0.8}{\volt} & \SI{0.8}{\volt}\\
    Frequency & \SI{0.5}{\giga\hertz} & \SI{1.4}{\giga\hertz} & \SI{1}{\giga\hertz} & \SI{0.8}{\giga\hertz} & \SI{1.23}{\giga\hertz} & \SI{1.26}{\giga\hertz}\\
    Measurement setup & silicon & silicon & post-layout & silicon & post-layout & post-layout \\
    \midrule
    Num Tiles & 25 (5 $\times$ 5 mesh) & 496 (8 $\times$ 62 mesh) & 256 (16 $\times$ 16 mesh) & 34 (6 $\times$ 6 mesh) & 1$^e$ & 32 (8 $\times$ 4 mesh) \\
    Die Area & \SI{36}{\square\milli\meter} & \SI{12.3}{\square\milli\meter} & n.A.$^e$ & \SI{64}{\square\milli\meter} & \SI{1.1}{\square\milli\meter} & \SI{39}{\square\milli\meter}\\
    NoC Area & \SI{0.84}{\square\milli\meter} & \SI{0.93}{\square\milli\meter} & n.A.$^e$ & n.A. & \SI{0.11}{\square\milli\meter} & \SI{1.37}{\square\milli\meter}\\
    NoC/Die Area Ratio & \textbf{\SI{2.9}{\percent}} &\SI{7.77}{\percent} & \SI{18.2}{\percent}$^c$/\SI{35.3}{\percent}$^d$ & n.A. & \SI{10}{\percent} & \SI{3.5}{\percent}\\
    \midrule
    Tile-to-Tile \gls{NOC} BW$^a$ & \SI{96}{\giga\bitpersecond} & \SI{45}{\giga\bitpersecond} & \SI{256}{\giga\bitpersecond}& \SI{310}{\giga\bitpersecond} & \SI{787}{\giga\bitpersecond} & \textbf{\SI{806}{\giga\bitpersecond}}\\
    Aggregate \gls{NOC} BW & \SI{4}{\tera\bitpersecond} & \textbf{\SI{361}{\tera\bitpersecond}} & n.A.$^e$ & \SI{74}{\tera\bitpersecond} & n.A.$^e$& \SI{103}{\tera\bitpersecond}\\
    \midrule
    Energy-efficiency & \SI{0.45}{\pico\joule\per\byte\per\hop} & n.A. & n.A. & \SI{2.0}{\pico\joule\per\byte\per\hop} & \SI{0.19}{\pico\joule\per\byte\per\hop} & \textbf{\SI{0.15}{\pico\joule\per\byte\per\hop}}\\
    NoC Power contrib. & n.A. & n.A. & n.A. & \SI{23}{\percent} & \SI{7}{\percent} & \textbf{\SI{5.7}{\percent}}\\
    \midrule
    Virt. Channels & \xmark & \xmark & \xmark & \xmark & \xmark & \xmark \\
    Multiple Phys. Channels & \cmark  (3) & \xmark & \xmark & \cmark (6) &  \cmark (3) & \cmark  (3) \\
    Link Data Width & 64 & 32 & 256 & 64 & 2$\times$64 + 512 & 2$\times$64 + 512\\
    \bottomrule
    \end{tabular}
    \label{tab:soa_comparison}
    \\
    $^a$simplex, counting only data-bits $^c$c4 dummy tile of size \SI{0.14}{\square\milli\meter} $^d$c1 dummy tile of size \SI{0.034}{\square\milli\meter} $^e$ only single tile\\
\end{table*}

\section{Conclusion}

In this paper, we presented an open-source \gls{NOC} design with full \gls{AXI} support tailored to handle the significant bandwidth demands of today's data-intensive applications. Our approach, which utilizes wide links, leverages advancements in modern technologies that make it physically feasible to accommodate an increased number of wires without the need for frequency multiplication.

Implemented in a \SI{12}{\nano\meter} \gls{VLSI} technology, an $8 \times 4$ compute mesh with 288 RISC-V cores has a low area overhead of just 3.5\% per compute tile. It delivers extremely high bandwidth, achieving up to \SI{645}{\giga\bitpersecond} per link and a total aggregate bandwidth of \SI{103}{\tera\bitpersecond}. The \gls{NOC}’s unique end-to-end ordering system, powered by a multi-stream capable \gls{DMA}, simplifies the \gls{NI} and eliminates inter-stream dependencies, enhancing scalability and efficiency. These improvements led to a 30\% reduction in area and a 47\% increase in \SI{}{\giga\flops\double} within the same floorplan compared to a traditional \gls{AXI}-based \gls{SOC}. Our design significantly outperforms current state-of-the-art \glspl{NOC}, offering up to three times the energy efficiency and more than double the tile-to-tile bandwidth.

\added{For future work, we aim to explore more complex topologies to evaluate the trade-off between performance benefits and the increased complexity in the physical design.}

\section*{Acknowledgment}

The authors thank Tobias Senti for his valuable contributions to this work.
We utilized ChatGPT by OpenAI, for enhancing text readability and grammar across this manuscript. The use of AI was limited to text improvement and did not contribute to original content or research findings.

\ifCLASSOPTIONcaptionsoff
  \newpage
\fi



%
\bibliographystyle{IEEEtran}
\bibliography{IEEEabrv, bibliography.bib}

%
\vskip -3\baselineskip plus -1fil
\begin{IEEEbiography}[{\includegraphics[width=1in,height=1.25in,clip,keepaspectratio]{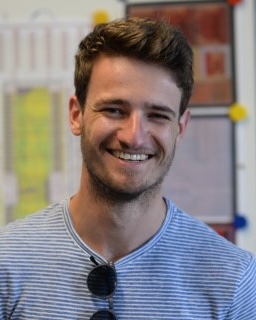}}]
{Tim Fischer} received his BSc and MSc in ``Electrical Engineering and Information Technology'' from the Swiss Federal Institute of Technology Zurich (ETHZ), Switzerland, in 2018 and 2021, respectively. He is currently pursuing a Ph.D. degree at ETH Zurich in the Digital Circuits and Systems group led by Prof. Luca Benini. His research interests include scalable and energy-efficient interconnects for both on-chip and off-chip communication.
\end{IEEEbiography}
\vskip -2.2\baselineskip plus -1fil

\begin{IEEEbiography}[{\includegraphics[width=1in,height=1.25in,clip,keepaspectratio]{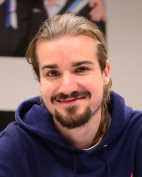}}]{Michael Rogenmoser}
received his BSc and MSc in ``Electrical Engineering and Information Technology'' from the Swiss Federal Institute of Technology Zurich (ETHZ), Switzerland, in 2020 and 2021, respectively. He is currently pursuing a Ph.D. degree in the Digital Circuits and Systems group of Prof. Benini. His research interests include fault-tolerant processing architectures and multicore heterogeneous SoCs for space.
\end{IEEEbiography}

\vskip -2.2\baselineskip plus -1fil
\begin{IEEEbiography}[{\includegraphics[width=1in,height=1.25in,clip,keepaspectratio]{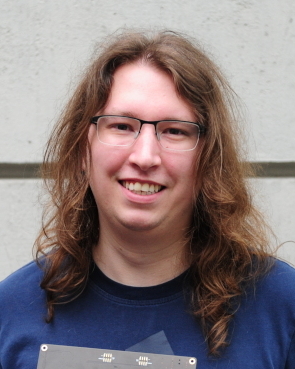}}]{Thomas Benz}
received the B.Sc. and M.Sc. degrees in electrical engineering and information technology from ETH Zurich, in 2018 and 2020, respectively. He is currently working toward the Ph.D. degree with the Digital Circuits and Systems group of Prof. Benini. His research interests include energy-efficient high-performance computer architectures and the design of ASICs.
\end{IEEEbiography}

\vskip -2.2\baselineskip plus -1fil
\begin{IEEEbiography}[{\includegraphics[width=1in,height=1.25in,clip,keepaspectratio]{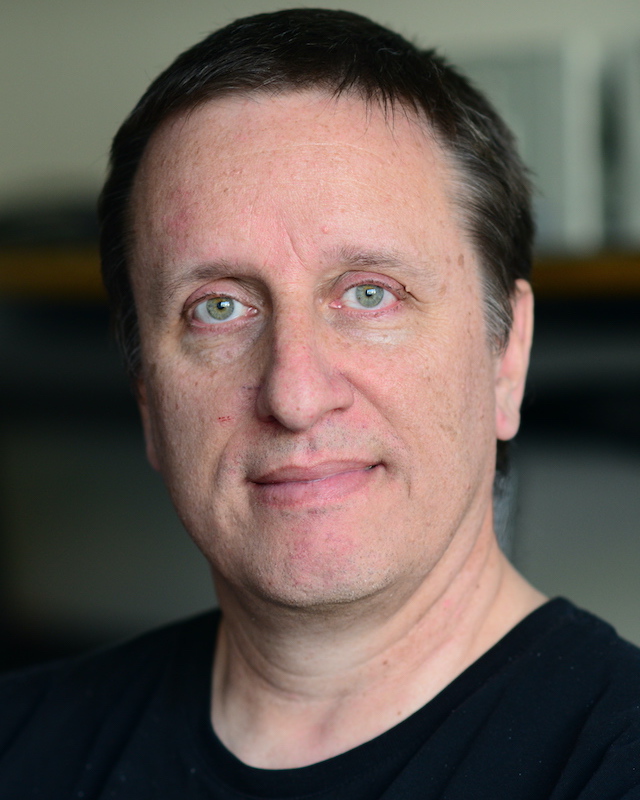}}]{Frank Gürkaynak}
received the BSc and MSc degrees in electrical engineering from the Istanbul Technical University, and the PhD degree in electrical engineering from ETH Zurich in 2006. He is currently working as a senior scientist with the Integrated Systems Laboratory, ETH Zurich. His research interests include digital low-power design and cryptographic hardware.
\end{IEEEbiography}

\vskip -2.2\baselineskip plus -1fil
\begin{IEEEbiography}[{\includegraphics[width=1.05in,height=1.25in,clip,keepaspectratio]{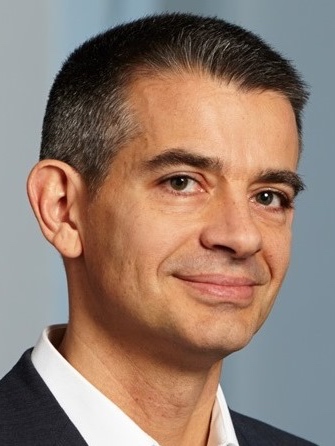}}]{Luca Benini}
holds the chair of digital Circuits and systems at ETHZ and is Full Professor at the Università di Bologna. He received a Ph.D. from Stanford University. Dr. Benini's research interests are in energy-efficient parallel computing systems, smart sensing micro-systems and machine learning hardware. He is a Fellow of the IEEE, of the ACM and a member of the Academia Europaea.
\end{IEEEbiography}





\end{document}